\documentclass[10pt]{article}
\usepackage{epsfig,amsmath,graphicx,amssymb,overpic,cite}

\usepackage{graphicx}
\usepackage{dcolumn}
\usepackage{bm}
\usepackage{graphicx}
\usepackage{subfigure}

\def\be{\begin{equation}}
\def\ee{\end{equation}}
\def\bee{\begin{eqnarray}}
\def\ene{\end{eqnarray}}
\def\bes{\begin{subequations}}
\def\ees{\end{subequations}}

\renewcommand {\thefootnote}{\dag}
\renewcommand {\thefootnote}{\ddag}
\renewcommand {\thefootnote}{ }

\begin{document}

\baselineskip=13pt
\renewcommand {\thefootnote}{\dag}
\renewcommand {\thefootnote}{\ddag}
\renewcommand {\thefootnote}{ }

\pagestyle{plain}

\begin{center}
{\Large \bf Higher-order rational solitons and rogue-like wave solutions of \\ the (2+1)-dimensional nonlinear fluid mechanics equations}
\end{center}

\begin{center}
Xiao-Yong Wen$^{a,b}$ and Zhenya Yan$^{a,*}$\footnote{$^{*}${\it Email address}: zyyan@mmrc.iss.ac.cn} \\[0.03in]
{\it \small $^{a}$Key Laboratory of Mathematics Mechanization, Institute
of Systems Science, AMSS, \\ Chinese Academy of Sciences, Beijing
100190, China \\
$^2$Department of Mathematics, School of Applied Science, Beijing Information
    Science and Technology University, Beijing  100192, China}
\end{center}



\vspace{0.2in}

\baselineskip=13pt

{\bf Abstract} \, The novel generalized perturbation $(n, M)$-fold Darboux transformations (DTs) are reported for the (2+1)-dimensional Kadomtsev-Petviashvili (KP) equation and its extension  by using the Taylor expansion of the Darboux matrix. The generalized perturbation $(1, N-1)$-fold DTs are used to find their higher-order rational solitons and rogue wave solutions in terms of determinants. The dynamics behaviors of these rogue waves are discussed in detail for different parameters and time, which display the interesting RW and soliton structures including the triangle, pentagon, heptagon profiles, etc. Moreover, we find that a new phenomenon that the parameter $(a)$ can control the wave structures of the KP equation from the higher-order rogue waves ($a\not=0$) into higher-order rational solitons ($a=0)$ in $(x,t)$-space with $y={\rm const}$.  These results may predict the corresponding dynamical phenomena in the models of fluid mechanics and other physically relevant systems.

\vspace{0.1in}\noindent {\bf Keywords} \, (2+1)-dimensional KP equation; Generalized perturbation Darboux transformation; rational solitons; rogue waves

\vspace{0.1in}

\section{Introduction}

Rogue waves (RWs), as a special phenomenon of solitary waves originally occurring in the deep ocean~\cite{org,org2,org3,org4}, have drawn more and more theoretical and experimental attention in many other fields such as nonlinear optics~\cite{orw,orw2,orw3}, hydrodynamics~\cite{hd}, Bose-Einstein condensates~\cite{bec1,bec2}, plasma physics~\cite{pla}, and finance~\cite{yan,yan2}. RWs are also known as freak waves~\cite{fw}, giant waves, great waves, killer waves, etc.
There is currently no unified concept for RWs, but RWs always possess two remarkable characteristics: on one hand, they are located in both space and time; on the other hand, they exhibit a smooth and dominant peak. They are always isolated huge waves with the amplitudes being two to three times than ones of its surrounding waves in the ocean~\cite{org}, they have caused many disastrous consequences in the ocean.

The focusing nonlinear Schr\"odinger (NLS) equation~\cite{nls,nls2,nls3,nls4,nls5}
\bee \label{nls1}
  iq_t+\frac12q_{xx}+|q|^2q=0,
 \ene
arising from many fields of nonlinear science such as nonlinear optics, the deep ocean, DNA, Bose-Einstein condensates, and finance, is an important model admitting the first-order RW solution in the rational form (also called Peregrine's RW solution)~\cite{ps}
\bee\label{ps}
q_{\rm ps}(x,t)=\displaystyle \left[1-\frac{4(1+2it)}{1+4(x^2+t^2)}\right]e^{it}
\ene
which can be regarded as the parameter limit of its breathers~\cite{MA,nail86,nail87,rgre}, and  higher-order RW solutions~\cite{nail,nail2,nail3}. The intensity $|q_{\rm ps}|^2$ is localized in both space and time, and approaches to one not zero as $x^2+t^2\rightarrow \infty$, as well as has three critical points, which differ from its bright soliton
\bee
 q_{\rm bs}(x,t)=\beta\, {\rm sech}(\beta x-v t)\exp\left[i\left(\frac{v}{\beta}x+\frac{\beta^4-v^2}{2\beta^2}t\right)\right],\quad
 \beta\not=0,\, v\in \mathbb{R}
 \ene
 in which $|q_{\rm bs}|^2\rightarrow 0$ as $|\beta x-v t|\rightarrow \infty$ and it has infinite critical points, that is a family of critical lines $\beta x-v t=0$. It has been shown that the Peregrine's RW solution has a good agreement with the numerical simulation and experimental results of Eq.~(\ref{nls1})~\cite{orw3}. Rogue waves were coined `rogons' if
 they reappear virtually unaffected in size or shape shortly after the interactions~\cite{yanpla10}.
 In fact, the singular rational solutions of the KdV equation were found by using some constructive methods such as the factorization of the Sturm-Liouville operators~\cite{kdv1,kdv2} and the limiting procedure on the solitons~\cite{kdv3}. As the (2+1)-D extension of the KdV equation, the KP-I equation
 \bee
 (u_{t}+6 uu_{x}+u_{xxx})_x-3u_{yy}=0,  \label{cmkdv}
\ene
was shown to admit the regular lower-order rational solutions (also called the lump solitons) in terms of the limit procedure of solitons~\cite{kdv3,kplum,kpm}. Moreover, the rational solutions of KP-I equation were also constructed from ones of the NLS equation~\cite{mat3,mat4}. But the KP-II equation given by Eq.~(\ref{cmkdv}) with $y\rightarrow \sqrt{-1}y$ was shown to admit the singular rational solutions~\cite{dt,mat2}.

Nowadays, to understand the physical mechanisms of RWs in nonlinear physical phenomena, some powerful methods studying the higher-order RW solutions of nonlinear wave equations have been drawn the increasing attention such as the modified and generalized Darboux transformation (DT)~\cite{nail,guo1,guo2, dt1,dt2a,dt3a}, the Hirota's bilinear method with the $\tau$-function~\cite{yang1,yang2}, the similarity transformation~\cite{yanpla10,yan12,yan12b, yann15}, etc. Among them, the DT method is an effective technique to solve the integrable nonlinear wave equations~\cite{dt82, dt75, dt,dt3,nail88,dt2}.

 It is well known that the DT can usually be used to study multi-soliton solutions of nonlinear integrable systems. In 2009, the usual DT can be used to study the multi-RW solutions of the focusing NLS equation (\ref{nls1}) with the plane-wave solution $q=e^{it}$~\cite{nail}. This modified DT method can theoretically be used to obtain the higher-order RW solutions of the focusing NLS equation (\ref{nls1}) but it is so complicated since the used DT can not give an explicit formulae for multi-RW solutions.  Recently, by means of the Taylor expansion and a limit procedure, the Matveev's generalized DT method~\cite{matveev92} was developed to construct the higher-order RW solutions of  the NLS and KN equations~\cite{guo1,guo2}. Furthermore, the RW solutions of several other nonlinear wave equations were also investigated using this method~\cite{zhq1}. Moreover recently, we presented a novel and simple method to find the generalized $(n, M)$-fold Darboux transformation of the modified NLS equation, the coupled AB system, and nonlocal NLS equation such that its higher-order rogue wave and rational soliton solutions were found using the  determinants~\cite{wen-yan,yanchao15,yanchao16}.

In this paper, we will present the new, generalized $N$-order DTs in terms of the Taylor series expansion and a limit procedure to directly obtain the higher-order rational solitons and RW solutions of the (2+1)-dimensional KP equation (\ref{cmkdv}) in the determinants. The main advantages of our method are that
no complicated iterations are used to obtain the higher-order RW solutions and the relations between the higher-order RW solutions and the `seed' solution are clear. Moreover, our method can also be applied to the (2+1)-D generalized  KP (gKP) equation~\cite{you}
\bee u_{t}+(1-2\kappa) uu_{x}-\kappa u_{xxx}-(1+\kappa) \partial^{-1}_x u_{yy}
+(\kappa+1/4)(u^{-1}u_x^2+u^{-1}\partial^{-1}_x u_{y})^2)_x=0.  \label{gkp}
\ene

The rest of this paper is organized as follows. In Sec. 2, we simply recall the usual DT of Eq.~(\ref{cmkdv}), and then present an idea to derive the generalized perturbation $(n, M)$-fold DT for the (2+1)-dimensional KP equation by using the Taylor expansion and a limit procedure such that its higher-order rational solitons and RW solutions in terms of   determinants, which contain the known results~\cite{kpm,kplum}. We also analyze their abundant wave structures. In Sec.3, we further extend this method to the (2+1)-D generalized KP equation (\ref{gkp}) such that its higher-order rational solitons and RW solutions can also be found. The method can also be extended to other higher-dimensional nonlinear wave equations. Some conclusions and discussions are given in the last section.

\section{Rational solitons and rogue waves of the (2+1)-D KP equation}

Kadomtsev and Petviashvili presented the (2+1)-dimensional KP equation (\ref{cmkdv}) (i.e., the (2+1)-dimensional extension of the KdV equation) when they relaxed the restriction that the waves was strictly one-dimensional. Eq.~(\ref{cmkdv}) is used to model the shallow-water waves with weakly nonlinear restoring forces and waves in ferromagnetic media. Eq.~(\ref{cmkdv}) can also describe the two-dimensional matter-wave pulses in Bose-Einstein condensates~\cite{huang03}. Moreover, the complete integrability of the KP hierarchy was shown in the sense of Frobenius~\cite{fkp}. By using a constraint, Eq.~(\ref{cmkdv}) can be decomposed into the focusing NLS equation and complex mKdV equation, and some new solutions of Eq.~(\ref{cmkdv}) have been obtained by using the usual DT~\cite{li, li2}. The KP equation can also be separated into a (1+1)-dimensional Broer-Kaup (BK) equation and a (1+1)-dimensional high-order BK equation by using the symmetry constraint~\cite{lou}. Based on the latter decomposition, a unified DT has been constructed such that some soliton-like solutions with five parameters were  obtained for the KP equation~\cite{fan1, fan2}. Moreover, the solutions of KP equation can also be expressed in terms of points on an infinite-dimensional Grassmannian~\cite{Will}.

 In the following, we firstly recall the $N$-fold DT of Eq.~(\ref{cmkdv}), and then we present the generalized DT such that we give its higher-order rational solitons and RW solutions. We  consider the following constraint
\bee
  u=2|p|^2, \label{yueshu}
\ene
where $p=p(x,y,t)$ is the complex function of the variables $x,y,t$. Thus, a decomposition of the KP equation (\ref{cmkdv}) is exactly related to the (1+1)-dimensional focusing NLS equation
\bee
 i p_{y} -p_{xx}-2|p|^2p=0 \label{nls}
\ene
and the (1+1)-D complex mKdV equation
\bee
  p_{t}+4p_{xxx}+24|p|^2p_x=0. \label{cmkdv2}
\ene
If $p(x,y,t)$ solves Eqs.~(\ref{nls}) and~(\ref{cmkdv2}), then the corresponding solution of the KP equation (\ref{cmkdv}) can be generated in terms of the constraint (\ref{yueshu}). In the following we mainly consider Eqs.~(\ref{nls}) and~(\ref{cmkdv2}).

\subsection{Lax pair and Darboux transformation}

The Lax representations (the linear iso-spectral problems) of Eqs.~(\ref{nls}) and~(\ref{cmkdv2}) are given as follows:
\bee
\hspace{-0.9in}\varphi_x=U \varphi, \quad U=\left(\begin{array}{cc}-i\lambda & ip \vspace{0.1in} \\  ip^{*}  & i\lambda \end{array}\right),  \label{lax1}
 \qquad
\ene
\bee
 \varphi_{y}=V \varphi,\quad V\!=\!\left(\!\begin{array}{cc}
        2 i\lambda^2-i|p|^2            & -2i \lambda p + p_x  \vspace{0.1in} \\
        -2i\lambda p^{*} -p^{*}_x      & -2 i\lambda^2+i|p|^2 \end{array} \!\right), \,\,\, \label{lax2}
\ene
\bee \label{lax3}
 \hspace{-0.9in} \varphi_{t}=W \varphi, \quad W=\left(\begin{array}{cc} W_{11} & W_{12} \vspace{0.1in}\cr W_{21} & W_{22}\end{array}\right) \quad
 \ene
with
 \bee\nonumber\begin{array}{l}
 W_{11}=-16 i \lambda^3+8 i \lambda |p|^2+4(pp^{*}_x-p_xp^{*}), \,\
 W_{12}= 16i \lambda^2p -8\lambda p_x  -4 i( p_{xx}+2|p|^2p), \vspace{0.1in} \\
W_{21}=16i\lambda^2p^{*} +8\lambda p^{*}_x-4i(p^{*}_{xx}+2|p|^2p^{*}),\,\,
W_{22}=16 i \lambda^3-8 i\lambda |p|^2-4(p p^{*}_x-p_xp^{*}),
\end{array}\ene
where the star represents the complex conjugation, $\varphi=(\phi,\psi)^T$ (the superscript $T$ denotes the vector transpose) is the vector eigenfunction, $\lambda$ is the spectral parameter,  and $i^2=-1$. It is easy to show that two zero curvature equations $U_y-V_x+[U, V]=0$ and $U_t-W_x+[U, W]=0$ yield Eqs.~(\ref{nls}) and~(\ref{cmkdv2}), respectively.

We consider the gauge transformation with the Darboux matrix $T(\lambda)$:
\bee\label{kpm}
  \widetilde{\varphi}=T(\lambda) \varphi, \qquad   \widetilde{\varphi}=( \widetilde{\phi},\,  \widetilde{\psi})^T \label{gauge}
\ene
which maps the old eigenfunction $\varphi$ into the new one $\widetilde{\varphi}$, where $\widetilde{\varphi}$ is required to satisfy
\bee \label{s21}
\widetilde{\varphi}_x=\widetilde{U} \widetilde{\varphi},\,\,  \widetilde{\varphi}_y=\widetilde{V} \widetilde{\varphi}, \,\,
\widetilde{\varphi}_{t}=\widetilde{W} \widetilde{\varphi},\,\,  \widetilde{U}=U|_{\{p=\widetilde{p},\, p^{*}=\widetilde{p}^{*}\}},  \,\
  \widetilde{V}=V|_{\{p=\widetilde{p},\, p^{*}=\widetilde{p}^{*}\}},  \,\,
    \widetilde{W}=W|_{\{p=\widetilde{p},\, p^{*}=\widetilde{p}^{*}\}}, \ene
where $T,\, U,\, \widetilde{U}, \, V, \widetilde{V},\, W$, and $\widetilde{W}$ satisfy
\bes\bee \label{kpt1}
T_{x}+[T,\,\, \{U,\,\, \widetilde{U}\}]=0, \vspace{0.1in} \\
\label{kpt2}
T_{y}+[T,\,\, \{V,\,\, \widetilde{V}\}]=0, \vspace{0.1in} \\
T_{t}+[T,\,\, \{W,\,\,  \widetilde{W}\}]=0,
\label{kpt3}\ene\ees
where we have introduced the generalized bracket for the square matrixes $[F, \{G,  \widetilde{G}\}]=FG-\widetilde{G}F$. Therefore we have
\bee\label{s2}
 \widetilde{U}_y-\widetilde{V}_x+[\widetilde{U},\widetilde{V}]=T(U_y-V_x+[U,\, V])T^{-1}\!=\!0,\quad\, \\
 \widetilde{U}_t-\widetilde{W}_x+[\widetilde{U},\widetilde{W}]=T(U_t-W_x+[U,\, W])T^{-1}\!=\!0, \quad\,
\ene
which yields the same equations (\ref{nls}) and (\ref{cmkdv2}) with $p\rightarrow \widetilde{p}$, i.e., $\widetilde{p}$ in the new spectral problem (\ref{s21}) is a solution of Eqs.~(\ref{nls}) and (\ref{cmkdv2}).

Hereby, based on Ref.~\cite{fan2}, the usual $N$-order Darboux matrix $T(\lambda)$ in Eq.~(\ref{kpm}) is chosen as
\bee \label{kpm}
 T(\lambda)=\left(\!\!\!\begin{array}{cc}
\lambda^{N}\!+\!\sum\limits_{j=0}^{N-1}A^{(j)}\lambda^{j}    &\sum\limits_{j=0}^{N-1}B^{(j)} \lambda^{j} \vspace{0.1in}\\
   -\!\sum\limits_{j=0}^{N-1}{B^{(j)}}^{*} \lambda^{j}   &\lambda^{N}\!+\!\sum\limits_{j=0}^{N-1}{A^{(j)}}^{*}\lambda^{j}\end{array}
       \!\!\!\right) \label{Tmatrix}, \qquad
\ene
where $N$ is a positive integer, $A^{(j)},B^{(j)}\, (0\leq j\leq N-1)$ are $2N$ unknown complex functions, which solve the linear algebraic system with $2N$ equations
\bee\label{kpsys}
\begin{array}{ll}
\left[\lambda_s^{N}\!+\!\sum\limits_{j=0}^{N-1}A^{(j)}\lambda_s^{j}\right]\phi_s(\lambda_s)\!+\!\sum\limits_{j=0}^{N-1}B^{(j)} \lambda_s^{j}\psi_s(\lambda_s)=0, \vspace{0.1in}\\
\left[\lambda_s^{N}\!+\!\sum\limits_{j=0}^{N-1}{A^{(j)}}^{*}\lambda_s^{j}\right]\psi_s(\lambda_s)\!-\!
\sum\limits_{j=0}^{N-1}{B^{(j)}}^{*} \lambda_s^{j}\phi_s(\lambda_s)=0, \end{array} \quad
\ene
where $\varphi_s(\lambda_s)=(\phi_s(\lambda_s),\psi(\lambda_s))^T\, (s=1,2,...,N)$ are the solutions of Lax pair (\ref{lax1})-(\ref{lax3}) for the spectral parameters $\lambda_s$ and the initial solution $p_0$. When $N$ distinct parameters $\lambda_j\ (\lambda_i\neq\lambda_j, i \neq j)$ are suitably chosen so that the determinant of the coefficients of $2N$ variables $A^{(j)},\, B^{(j)}\, (j=0,1,...,N-1)$ in system~(\ref{kpsys}) is nonzero, the Darboux matrix $T$ is uniquely determined by system (\ref{kpsys}).

It follows from system (\ref{kpsys}) that $2N$ distinct parameters $\lambda_k,\, \lambda_k^{*}\,$ $(\lambda_i\neq\lambda_j, \lambda_i\neq 0, i\neq j, i=1,2,...,N)$ are the roots of the $2N$-th order polynomial $\det T(\lambda)$, i.e.,
\bee
 \det T(\lambda)= \prod\limits_{k=1}^{N}(\lambda-\lambda_k)(\lambda-\lambda_k^{*})=0. \label{ab1}
\ene

To make sure that Eqs.~(\ref{kpt1})-(\ref{kpt3}) hold for the given Darboux matrix (\ref{Tmatrix}), the following Darboux transformation of Eq.~(\ref{cmkdv}) holds~\cite{fan2}. \\

\noindent{\bf Theorem 1.} {\it Let $\varphi_i(\lambda_i)=(\phi_i(\lambda_i),\psi_i(\lambda_i))^T\, (i=1,2,...,N)$ be $N$ distinct column vector solutions of the spectral problem (\ref{lax1})-(\ref{lax3}) for the  corresponding $N$ distinct spectral parameters $\lambda_1,\, \lambda_2,...,\lambda_N$ and the initial solution $p_0$ of Eqs.~(\ref{nls}) and (\ref{cmkdv2}), respectively, then the $N$-fold Darboux transformation of Eq.~(\ref{cmkdv}) is given by
 \bee\label{kpsol}
 \widetilde{u}_{N-1}=2|\widetilde{p}_{N-1}|^2=2\left|p_0+2B^{(N-1)}\right|^2,
\ene
where $B^{(N-1)}$ is given by $B^{(N-1)}=\frac{\Delta B^{(N-1)}}{\Delta_N}$ (cf.  system~(\ref{kpsys})) with
\bee\nonumber
\Delta_N = \left|\begin{array}{cccccccc}
\lambda_1^{N-1}\phi_1 & {\lambda_1}^{N-2}\phi_1  &\ldots \quad\quad & \phi_1 & {\lambda_1}^{N-1}\psi_1 &  \lambda_1^{N-2}\psi_1  &\ldots & \psi_1 \vspace{0.1in} \\
 \ldots & \ldots      &\ldots  \quad\quad          &\ldots &\ldots & \ldots      &\ldots           &\ldots \vspace{0.1in}  \\
\lambda_N^{N-1}\phi_N & {\lambda_N}^{N-2}\phi_N  &\ldots \quad\quad  & \phi_N & {\lambda_N}^{N-1}\psi_N & \lambda_N^{N-2}\psi_N  &\ldots & \psi_N \vspace{0.1in} \\
\lambda_1^{*(N-1)}\psi^{*}_1 & {\lambda_1}^{*(N-2)}\psi_1^{*}  &\ldots \quad\quad  & \psi_1^{*} & -\lambda_1^{*(N-1)}\phi_1 & -\lambda_1^{*(N-2)}\phi_1^{*}
 &\ldots & -\phi_1^{*} \vspace{0.1in}  \\
\lambda_2^{*(N-1)}\psi^{*}_2 & {\lambda_2}^{*(N-2)}\psi_2^{*}  &\ldots\quad\quad  & \psi_2^{*} & -\lambda_2^{*(N-1)}\phi_2 & -\lambda_2^{*(N-2)}\phi_2^{*}
&\ldots & -\phi_2^{*} \vspace{0.1in}  \\
\ldots & \ldots      &\ldots \quad\quad           &\ldots &\ldots & \ldots      &\ldots           &\ldots   \vspace{0.1in}  \\
\lambda_N^{*(N-1)}\psi^{*}_N & {\lambda_N}^{*(N-2)}\psi_N^{*}  &\ldots \quad\quad  & \psi_N^{*} & -\lambda_N^{*(N-1)}\phi_N & -\lambda_N^{*(N-2)}\phi_N^{*}
 &\ldots & -\phi_N^{*} \vspace{0.1in}  \\
\end{array}\right|,
\ene
and $\Delta B^{(N-1)}$ is given by the determinant $\Delta_N$ by replacing its $(N+1)$-th column by the column vector $(-\lambda_1^N\phi_1$, $-\lambda_2^N\phi_2$,$\cdots$, $-\lambda_N^N\phi_N$, $-\lambda_1^{N*}\psi_1^{*}$, $-\lambda_2^{N*}\psi_2^{*}$,$\cdots$, $-\lambda_N^{N*}\psi_N^{*})^T$.
} \\

By applying the $N$-fold DT (\ref{kpsol}), the multi-soliton solutions for Eq.~(\ref{cmkdv}) have been obtained by choosing constant seed solution (e.g., $p_0=c$)~\cite{fan2}.
 
 In the following, we will construct the generalized perturbation $(n, M)$-fold DT and higher-order RW solutions in terms of determinant through the $N$-order Darboux matrix $T$, the Taylor expansion and a limit procedure.

\subsection{Generalized perturbation $(1, N-1)$-fold Darboux transformation}

To study other types of solutions of Eq.~(\ref{cmkdv}) such as multi-rogue wave solutions, we need to change some functions $A^{(j)}$ and $B^{(j)}$ in the above-mentioned Darboux matrix $T$ given by Eq.~(\ref{kpm}) and the initial solution $q_0$ such that we may obtain other types of solutions of Eq.~(\ref{cmkdv}) in terms of some generalized DTs.

Here we still consider the Darboux matrix (\ref{kpm}), but we only consider one spectral parameter $\lambda=\lambda_1$ not $N$ spectral parameters $\lambda=\lambda_k\, (k=1,2,...,N)$, in which  the condition $T(\lambda_1)\varphi(\lambda_1)=0$ leads to the system
 \bee\label{nlsag1}
 \left[\lambda_1^{N}\!+\!\sum\limits_{j=0}^{N-1}A^{(j)}\lambda_1^{j}\right]\!\phi(\lambda_1)
  +\sum\limits_{j=0}^{N-1}B^{(j)}\lambda^{j}_1\psi(\lambda_1)=0,\qquad\quad \\
\label{nlsbg2}
 \left[\lambda_1^{N*}\!+\!\sum\limits_{j=0}^{N-1}\!{A^{(j)}}\lambda_1^{j*}\right]\!\psi^{*}(\lambda_1)\!-\!\sum\limits_{j=0}^{N-1}{B^{(j)}} \lambda_1^{j*}\phi^{*}(\lambda_1)\!=\!0,\qquad
\ene
where $(\phi(\lambda_1), \psi(\lambda_1))^T$ is a solution of the linear spectral problem (\ref{lax1})-(\ref{lax3}) with the spectral parameter $\lambda=\lambda_1$.

 These two linear algebraic Eqs.~(\ref{nlsag1}) and (\ref{nlsbg2}) contain the $2N$ unknown functions $A^{(j)}$ and $B^{(j)}\, (j=0,1,...,N-1)$. If $N=1$, then we can determine only two complex functions $A^{(0)}$ and $B^{(0)}$ from Eqs.~(\ref{nlsag1}) and (\ref{nlsbg2}) in which we can not deduce the different functions $A^{(0)}$ and $B^{(0)}$ comparing from the above-mentioned DT; If $N>1$, then we have $2(N-1)>2$ free functions for $A^{(j)}$ and $B^{(j)}\, (j=0,1,...,N-1)$. This means that the number of the unknown variables $A^{(j)}$ and $B^{(j)}$ is larger than one of equations such that we have some free functions, which seems to be useful for the Darboux matrix, but it may be difficult to show the invariant conditions (\ref{kpt1})-(\ref{kpt3}).

  To determine more (e.g., $2(N-1)$) constraint equations for the complex functions $A^{(j)}$ and $B^{(j)}\, (j=0,1,...,N)$ except for the given constraints (\ref{nlsag1}) and (\ref{nlsbg2}), we need to expand the expression $T(\lambda_1)\varphi(\lambda_1)\big|_{\lambda_1=\lambda_1+\varepsilon}=T(\lambda_1+\varepsilon)\varphi(\lambda_1+\varepsilon)$  at $\varepsilon=0$.
We know that \bee \label{nlsp-p}
 \varphi(\lambda_1+\varepsilon)\!=\!\varphi^{(0)}(\lambda_1)\!+\!\varphi^{(1)}(\lambda_1)\varepsilon
 \!+\!\varphi^{(2)}(\lambda_1)\varepsilon^2\!+\!\cdots,\quad
\ene
where $\varphi^{(k)}(\lambda_1)=\frac{1}{k!}\frac{\partial^k}{\partial \lambda^k}\varphi(\lambda)|_{\lambda=\lambda_1}$ and
\bee \label{nlsme-p}
T(\lambda_1+\varepsilon)=\sum\limits_{k=0}^{\infty} T^{(k)}(\lambda_1)\varepsilon^k, \label{zhan1}
\ene
where $T^{(0)}(\lambda_1)=T(\lambda_1),\,\, T^{(N)}(\lambda_1)=I$, and
\bee\label{td} 
\nonumber
T^{(k)}(\lambda_1)=\left(\begin{array}{cc}
C^{k}_N \lambda_1^{N\!-\!k}\!\!+\!\!\sum\limits_{j=k}^{N-1}\!C^{k}_j A^{(j)}\lambda_1^{j-k}   & \sum\limits_{j=k}^{N-1}C^{k}_j B^{(j)}\lambda_1^{j-k} \vspace{0.1in} \\
-\sum\limits_{j=k}^{N-1}C^{k}_j {B^{(j)}}^{*}\lambda_1^{j-k}                   & C^{k}_N \lambda_1^{N\!-\!k}\!\!+\!\!\sum\limits_{j=k}^{N-1}\!\!C^{k}_j {A^{(j)}}^{*}\lambda_1^{j-k}\\ \end{array} \!\!\right) \ene
with $C^{k}_j=j (j-1)...(j-k+1)/k!$\, ($k=0,1,...,N$).

Therefore, it follows from Eqs.~(\ref{nlsp-p}) and (\ref{nlsme-p}) that we obtain
\bee\label{tp-p}
 T(\lambda_1+\varepsilon) \varphi(\lambda_1+\varepsilon)
  \!\!=\!\!\sum_{k=0}^{\infty}\sum\limits_{j=0}^{k}T^{(j)}(\lambda_1)\varphi^{(k-j)}(\lambda_1)\varepsilon^k. \quad
\ene
To determine the $2N$ unknown functions $A^{(j)},\, B^{(j)}\,$ $(0\leq j\leq N-1)$ in Eq.~(\ref{kpm}), let
 \bee\nonumber
\lim\limits_{\varepsilon \to 0}\dfrac{T(\lambda_1+\varepsilon) \varphi(\lambda_1+\varepsilon)}{\varepsilon^k}=0,\,\,\,
(k=0,1,...,N-1). \qquad \label{jixian1-p}
\ene
Then, we obtain the linear algebraic system with the $2N$ equations
\bee\label{nls1sys-p}\begin{array}{r}
T^{(0)}(\lambda_1)\varphi^{(0)}(\lambda_1)=0,  \vspace{0.1in}\\
T^{(0)}(\lambda_1)\varphi^{(1)}(\lambda_1)+T^{(1)}(\lambda_1)\varphi^{(0)}(\lambda_1)=0, \vspace{0.1in} \\
\qquad \cdots\cdots,\qquad\qquad  \vspace{0.1in} \\
\sum\limits_{j=0}^{N-1}T^{(j)}(\lambda_1)\varphi^{(N-1-j)}(\lambda_1)=0,
\end{array}\ene
in which the first matrix system, $T^{(0)}(\lambda_1)\varphi^{(0)}(\lambda_1)=T(\lambda_1)\varphi(\lambda_1)=0$, is just Eqs.~(\ref{nlsag1}) and (\ref{nlsbg2}).

Therefore we have introduced system (\ref{nls1sys-p}) containing $2N$ algebraic equations  with $2N$ unknowns functions $A^{(j)}$ and $B^{(j)}\, (j=0,1,...,N-1)$.
When the eigenvalue $\lambda_1$ is suitably chosen so that the determinant of the coefficients for system~(\ref{nls1sys-p}) is nonzero, hence the transformation matrix $T$ can be uniquely determined by system~(\ref{nls1sys-p}). Owing to new distinct functions $A^{(j)}, B^{(j)}$ obtained in the $N$-order Darboux matrix $T$, so we can derive the `new' DT with the same eigenvalue $\lambda=\lambda_1$.

\vspace{0.1in}
\noindent {\bf Theorem 2.} {\it Let $\varphi(\lambda_1)=(\phi(\lambda_1),\psi(\lambda_1))^T$ be a column vector solution of the spectral problem (\ref{lax1})-(\ref{lax3}) for the spectral parameter $\lambda_1$ and initial solution $q_0$ of Eqs.~(\ref{nls}) and (\ref{cmkdv2}), then the generalized perturbation $(1,N-1)$-fold Darboux transformation of Eq.~(\ref{cmkdv}) is given by
\bee\label{kpsol-p}
  \widetilde{u}_{N-1}=2|\widetilde{p}_{N-1}|^2=2\left|p_0+2B^{(N-1)}\right|^2,
\ene
where $B^{(N-1)}$  is given by $B^{(N-1)}=\frac{\Delta B^{(N-1)}}{\Delta_{N}}$ (cf. system (\ref{nls1sys-p})) with
$\Delta_{N}={\rm det}[(\Delta_{j,s})_{2N\times 2N}]=$
\bee \nonumber\left|\begin{array}{llllllll}
{\lambda_1}^{(N-1)} {\phi^{(0)}} & {\lambda_1}^{(N-2)} {\phi^{(0)}}  &\ldots & {\phi^{(0)}} & {\lambda_1}^{(N-1)} {\psi^{(0)}} & {\lambda_1}^{(N-2)} {\psi^{(0)}}  &\ldots & {\psi^{(0)}} \vspace{0.1in} \\
\Delta_{2,1}& \Delta_{2,2}  &\ldots &{\phi^{(1)}}  & \Delta_{2,N+1} &  \Delta_{2,N+2} &\ldots &{\psi^{(1)}} \vspace{0.1in}\\
 \ldots & \ldots      &\ldots           &\ldots &\ldots & \ldots      &\ldots           &\ldots  \vspace{0.1in} \\
\Delta_{N,1}  & \Delta_{N,2} &\ldots &{\phi^{(N-1)}} &\Delta_{N,N+1}  & \Delta_{N,N+2}  &\ldots &{\psi^{(N-1)}} \vspace{0.1in}\\
{\lambda_1}^{*(N-1)} {\psi^{(0)*}} & {\lambda_1}^{*(N-2)} {\psi^{(0)*}}  &\ldots & {\psi^{(0)*}} & -{\lambda_1}^{*(N-1)} {\phi^{(0)*}} & -{\lambda_1}^{*(N-2)} {\phi^{(0)*}} &\ldots & -{\phi^{(0)*}}\vspace{0.1in} \\
\Delta_{N+2,1} & \Delta_{N+2,2}  &\ldots &{\psi^{(1)*}}  &\Delta_{N+2,N+1} & \Delta_{N+2,N+2}  &\ldots &-{\phi^{(1)*}} \vspace{0.1in}\\
 \ldots & \ldots      &\ldots           &\ldots &\ldots & \ldots      &\ldots           &\ldots  \\
\Delta_{2N,1}  & \Delta_{2N,2}  &\ldots &{\psi^{*(N-1)}} &\Delta_{2N,N+1} & \Delta_{2N,N+2}  &\ldots &-{\phi^{(N-1)*}} \\
\end{array}\right|,
\ene
where $\Delta_{j,s}\, (1\leq j, s\leq 2N$) are given by the following formulae:
\bee\nonumber
\Delta_{j,s}=\begin{cases}\sum\limits_{k=0}^{j-1}C^{k}_{N-s} {\lambda_1}^{(N-s-k)} {\phi^{(j-1-k)}}
               \quad {\rm for} \quad 1\leq j,\, s\leq N, \vspace{0.1in} \\
 \sum\limits_{k=0}^{j-1}C^{k}_{2N-s} {\lambda_1}^{(2N-s-k)} {\psi^{(j-1-k)}} \quad {\rm for} \quad 1\leq j\leq N,\, N+1\leq s\leq 2N, \vspace{0.1in}\\
\sum\limits_{k=0}^{j-(N+1)}C^{k}_{N-s} {\lambda_1}^{*(N-s-k)} {\psi^{(j-N-1-k)*}} \quad {\rm for} \quad
N+1\leq j\leq 2N,\, 1\leq s\leq N, \vspace{0.1in} \\
-\sum\limits_{k=0}^{j-(N+1)}C^{k}_{2N-s} {\lambda_1}^{*(2N-s-k)} {\phi^{(j-N-1-k)*}} \quad {\rm for} \quad
N+1\leq j,\, s \leq 2N \end{cases} \ene
and $\Delta {B^{(N-1)}}$ is found from $\Delta_N$ by replacing its
$(N+1)$-th column with the  vector $b=(b_j)_{2N\times 1}$, where
\bee\nonumber
b_j=\begin{cases} -\sum\limits_{k=0}^{j-1}C^{k}_{N} {\lambda_1}^{(N-k)} {\phi^{(j-1-k)}}  \quad {\rm for} \quad  1\leq j\leq N  \vspace{0.1in} \\ -\sum\limits_{k=0}^{j-(N+1)}C^{k}_{N} {\lambda_1}^{*(N-k)} {\psi^{(j-N-1-k)*}}  \quad {\rm for} \quad  N+1\leq j\leq 2N.
\end{cases} \ene
}

\subsection{Generalized perturbation $(n, M)$-fold Darboux transformation}

Here we consider the Darboux matrix (\ref{Tmatrix}) and assume that the eigenfunctions $\varphi_i(\lambda_i)\, (i=1,2,...,n)$ are the solutions of the linear spectral problem (\ref{kpt1})-(\ref{kpt3}) for the spectral parameter $\lambda_i$ and initial solution $p_0$ of Eqs.~(\ref{nls}) and (\ref{cmkdv2}). Thus we have
\bee\label{kpg}
 T(\lambda_i+\varepsilon) \varphi_i(\lambda_i+\varepsilon)
  \!\!=\!\!\sum_{k=0}^{\infty}\sum\limits_{j=0}^{k}T^{(j)}(\lambda_i)\varphi_i^{(k-j)}(\lambda_i)\varepsilon^k, \quad\,\,
  i=1,2,...,n,
\ene
where $\varphi_i^{(k)}(\lambda_i)=\frac{1}{k!}\frac{\partial^k}{\partial \lambda_i^k}\varphi_i(\lambda_i)$, $T^{(j)}(\lambda_i)$  ($j=0,1,2,...)$ are similar to Eq.~(\ref{td}), and $\varepsilon$ is a small parameter.

It follows from Eq.~(\ref{kpg}) and
 \bee\nonumber
\lim\limits_{\varepsilon \to 0}\dfrac{T(\lambda_i+\varepsilon) \varphi_i(\lambda_i+\varepsilon)}{\varepsilon^{k_i}}=0
 \label{jixiankp}
\ene
with $i=1,2,...,n$ and $k_i=0,1,...,m_i$ that we obtain the linear algebraic system with the $2N$ equations ($N=n+\sum_{i=1}^nm_i=n+M$):
\bee\label{kpsysg}\begin{array}{r}
T^{(0)}(\lambda_i)\varphi_i^{(0)}(\lambda_i)=0,  \vspace{0.1in}\\
T^{(0)}(\lambda_i)\varphi_i^{(1)}(\lambda_i)+T^{(1)}(\lambda_i)\varphi_i^{(0)}(\lambda_i)=0, \vspace{0.1in} \\
\qquad \cdots\cdots,\qquad\qquad  \vspace{0.1in} \\
\sum\limits_{j=0}^{m_i}T^{(j)}(\lambda_i)\varphi_i^{(m_i-j)}(\lambda_i)=0,
\end{array}\ene
$i=1,2,...,n$, in which we have first several systems for every index $i$, i.e., $T^{(0)}(\lambda_i)\varphi_i^{(0)}(\lambda_i)=T(\lambda_i)\varphi_i(\lambda_i)=0$ are just some ones in system~(\ref{kpsys}), but they are different if there exist at least one index $m_i\not=0$. \\

\noindent {\bf Theorem 3.}\, {\it Let $\varphi_i(\lambda_i)=(\phi_i(\lambda_i),\psi_i(\lambda_i))^T\,\, (i=1,2,...,n)$ be column vector solutions of the spectral problem (\ref{lax1})-(\ref{lax3}) for the $n$ spectral parameters $\lambda_1, \lambda_2,...,\lambda_n$ and initial solution $p_0$ of Eqs.~(\ref{nls}) and (\ref{cmkdv2}), respectively, then the generalized perturbation $(n, M)$-fold DT of Eq.~(\ref{cmkdv}) is given by
\bee\label{kpsolg}
 \widetilde{u}_{N-1}=2|\widetilde{p}_{N-1}|^2=2\left|p_0+2B^{(N-1)}\right|^2,
\ene
where $B^{(N-1)}$ is given by $B^{(N-1)}=\frac{\Delta B^{(N-1)}}{\Delta_N^{(n)}}$  (cf. system ~(\ref{kpsysg})) with
$\Delta_N^{(n)}={\rm det}([\Delta^{(1)}...\Delta^{(n)}]^T)$ and $\Delta^{(i)}\!=\!(\Delta^{(i)}_{j,s})_{2N\times 2N}=$
\bee\nonumber
\left[\begin{array}{llllllll}
{\lambda_i}^{N-1} {\phi_i^{(0)}} & {\lambda_i}^{N-2} {\phi_i^{(0)}}  &\ldots & {\phi_i^{(0)}} & {\lambda_i}^{N-1} {\psi_i^{(0)}} & {\lambda_i}^{N-2} {\psi_i^{(0)}}  &\ldots & {\psi_i^{(0)}} \vspace{0.1in} \\
\Delta_{2,1}^{(i)}& \Delta_{2,2}^{(i)}  &\ldots &{\phi_i^{(1)}}  & \Delta_{2,N+1}^{(i)} &  \Delta_{2,N+2}^{(i)} &\ldots &{\psi_i^{(1)}} \vspace{0.1in}\\
 \ldots & \ldots      &\ldots           &\ldots &\ldots & \ldots      &\ldots           &\ldots  \vspace{0.1in} \\
\Delta_{m_i+1,1}^{(i)}  & \Delta_{m_i+1,2}^{(i)} &\ldots &{\phi_i^{(m_i)}} &\Delta_{m_i+1,N+1}^{(i)}  & \Delta_{m_i+1,N+2}^{(i)}  &\ldots &{\psi_i^{(m_i)}} \vspace{0.1in}\\
{\lambda_i^*}^{(N-1)} {\psi_i^{(0)}}^* & {\lambda_i^*}^{(N-2)} {\psi_i^{(0)}}^*  &\ldots & {\psi_i^{(0)}}^* & -{\lambda_i^*}^{(N-1)} {\phi_i^{(0)}}^* & -{\lambda_i^*}^{(N-2)} {\phi_i^{(0)}}^*  &\ldots & -{\phi_i^{(0)}}^*\vspace{0.1in} \\
\Delta_{m_i+3,1}^{(i)} & \Delta_{m_i+3,2}^{(i)}  &\ldots &{\psi_i^{(1)}}^*  &\Delta_{m_i+3,N+1}^{(i)} & \Delta_{m_i+3,N+2}^{(i)}  &\ldots &-{\phi_i^{(1)}}^* \vspace{0.1in}\\
 \ldots & \ldots      &\ldots           &\ldots &\ldots & \ldots      &\ldots           &\ldots  \\
\Delta_{2(m_i+1),1}^{(i)}  & \Delta_{2(m_i+1),2}^{(i)}  &\ldots &{\psi_i^{(m_i)}}^* &\Delta_{2(m_i+1),N+1}^{(i)} & \Delta_{2(m_i+1),N+2}^{(i)}  &\ldots &-{\phi_i^{(m_i)}}^* \\
\end{array}\right],
\ene
where $\Delta_{j,s}^{(i)}\, (1\leq j, s\leq 2N$) are given by the following formulae:
\bee\nonumber
\Delta_{j,s}^{(i)}=\begin{cases}
 \sum\limits_{k=0}^{j-1}C^{k}_{N-s} {\lambda_i}^{N-s-k} {\phi_i^{(j-1-k)}}
\quad {\rm for} \quad 1\leq j\leq m_i+1,\, 1\leq s\leq N, \vspace{0.1in} \\
\sum\limits_{k=0}^{j-1}C^{k}_{2N-s} {\lambda_i}^{2N-s-k} {\psi_i^{(j-1-k)}}\quad {\rm for} \quad 1\leq j\leq m_i+1,\, N+1\leq s\leq 2N, \vspace{0.1in}\\
\sum\limits_{k=0}^{j-(N+1)}C^{k}_{N-s} {\lambda_i}^{*(N-s-k)} {\psi_i^{(j-N-1-k)*}} \quad {\rm for} \quad
m_i+2\leq j\leq 2(m_i+1),\, 1\leq s\leq N, \vspace{0.1in}\\
-\sum\limits_{k=0}^{j-(N+1)}C^{k}_{2N-s} {\lambda_i}^{*(2N-s-k)} {\phi_i^{(j-N-1-k)*}}\quad {\rm for} \quad
m_i+2\leq j\leq 2(m_i+1),\, N+1\leq s\leq 2N \end{cases} \ene
and $\Delta {B^{(N-1)}}$ is obtained from the determinant $\Delta_N^{(n)}$ by replacing its
$(N+1)$-th column with the vector $(b^{(1)}\cdots b^{(n)})^T$, where $b^{(i)}=(b_j^{(i)})_{2(m_i+1)\times 1}$ with
\bee\nonumber
b_j^{(i)}=\begin{cases}
-\sum\limits_{k=0}^{j-1}C^{k}_{N} {\lambda_i}^{N-k} {\phi_i^{(j-1-k)}} \quad {\rm for} \quad  1\leq j\leq m_i+1, \vspace{0.1in}  \\ -\sum\limits_{k=0}^{j-(N+1)}C^{k}_{N} {\lambda_i}^{*(N-k)} {\psi_i^{(j-N-1-k)*}} \quad {\rm for} \quad m_i+2\leq j\leq 2(m_i+1).
\end{cases} \ene
}

 Since the new distinct functions $A^{(j)}$ and $B^{(j)}$ obtained in the $N$-order Darboux matrix $T$ by solving system (\ref{kpsysg}) with at least one $m_i\not=0$, which are different from system ~(\ref{kpsys}) generating the functions $A^{(j)}$ and $B^{(j)}$ in the usual DT transformation, thus we may deduce the new $(1,N-1)$-fold DT with the same eigenvalue $\lambda=\lambda_1$ or the $(n,M)$-fold DT with the $n$ distinct eigenvalues $\lambda_i\, (i=1,2,...,n)$ such that the new solutions may be derived. We call Eqs.~(\ref{gauge}) and (\ref{kpsolg}) associated with new $A^{(j)}, B^{(j)}$ defined by system (\ref{kpsysg}) as a generalized perturbation $(n, M)$-fold DT of Eq.~(\ref{cmkdv}).\\

 {\bf Remark.} System~(\ref{kpsysg}) in the generalized perturbation $(n, M)$-fold DT is very important. System (\ref{kpsysg}) is similar to system~(\ref{kpsys}) in the usual $N$-fold DT and both of them can determine the unknown functions $A^{(j)}$ and $B^{(j)}$ of the Darboux matrix (\ref{kpm}) in the gauge transformation, but they are different from each other: for the unknown Darboux matrix $T$ (\ref{kpm}),  System~(\ref{kpsys}) contains the $N$ distinct eigenvalues, while system~(\ref{kpsysg}) has at most $N$ eigenvalues. Because of the different $A^{(j)}$ and $B^{(j)}$, the former can lead to the multi-soliton solutions, while the latter for the case $n<N$ may generate new solutions, e.g., the higher-order rational solutions including higher-order RW solutions. \\

Before we consider the higher-order rational solitons and RW solutions of the KP equation (\ref{cmkdv}), we give the following proposition, which is useful to generate its new solutions in terms of the known solutions. \\

\noindent {\bf Proposition 1.} {\it If $u(x,y,t)$ is a solution of the KP eqaution (\ref{cmkdv}), then so is \bee
\hat{u}(x,y,t)=\alpha^2u(\alpha x+\beta y+\gamma t, \alpha^2y+6\alpha\beta t, \alpha^3t)+\frac{3\beta^2-\alpha \gamma}{6\alpha^2},\ene
 where $\alpha\not=0,\,\beta,\,\gamma$ are real-valued constants}.\\

Therefore, for the obtained solution $\widetilde{u}_{N-1}$ given by Eq.~(\ref{kpsolg}) due to the generalized perturbation $(n, M)$-fold DT, we can further generate the `new' solution of the KP equation (\ref{cmkdv}) by using Proposition 1.

\subsection{The higher-order rational solitons and rogue wave solutions}

In what follows, we shall present some higher-order rational solitons and RW solutions of Eq.~(\ref{cmkdv}) in terms of determinants using the generalized perturbation $(n, M)$-fold DT. we will use a plane wave solution as a seed solution of Eq.~(\ref{cmkdv}) and consider only one spectral parameter $\lambda=\lambda_1$, i.e., $n=1,\, m_1=N-1$ in Theorem 3.

We begin with the non-trivial `seed' plane wave solution of Eqs.~(\ref{nls}) and (\ref{cmkdv2})
 \bee \label{is}
  p_0(x,y,t)=e^{i[ax+(a^2-2)y+(4a^3-24 a)t]},
   \ene
  which differs from the chosen zero seed solution to study multi-soliton solutions~\cite{kplum}, where $a$ is a real parameter, the wave numbers in $x-$ and $y$-directions are $a$ and $a^2-2$, respectively. It is known that the phase velocities in $x-$ and $y$-directions are $4(6-a^2)$ and $\frac{4a(6-a^2)}{a^2-2}$, respectively, and the group velocities in $x-$ and $y$-directions are $12(2-a^2)$ and $12-6a^2$, respectively.

Substituting Eq.~(\ref{is}) into the spectral problem (\ref{lax1})-(\ref{lax3}) leads to the eigenfunction solution of the Lax pair (\ref{lax1})-(\ref{lax3}) as follows:
\bee
\varphi=\left[\begin{array}{c}(C_1 e^{Q}+C_2 e^{-Q})e^{\frac{i}{2}[ax+(a^2-2)y+(4a^3-24a)t]} \vspace{0.1in}\\ (C_2 e^{Q}-C_1 e^{-Q})e^{-\frac{i}{2}[ax+(a^2-2)y+(4a^3-24a)t]} \end{array} \right],
\label{e29}
\ene
with
\bee\nonumber \begin{array}{l}
C_1=C_-,\,\, C_2=C_+,\,\,
C_{\pm}=\sqrt{\dfrac{\sqrt{[a+2\lambda)^2+4]}}{2}
  \pm \dfrac{(8\lambda^3-4\lambda+a^3-2a)}{2(4\lambda^2-2a\lambda+a^2-2)}}, \vspace{0.1in}\\
  Q=\dfrac{i}{2}\sqrt{(a+2\lambda)^2+4}[x+(a-2\lambda)y +4(4\lambda^2-2a\lambda+a^2-2)t+\Theta(\varepsilon)], \,\, \Theta(\varepsilon)=\sum\limits_{k=1}^{N}(b_k+ic_k)\varepsilon^{2k},
\end{array}
 \ene
where $b_k,c_k (k=1,2,...,N)$ are real parameters and $\varepsilon$ is a small parameter.

Next, we firstly fix the eigenvalue $\lambda_{1}=i-\frac{a}{2}$ and set $\lambda=\lambda_{1}+\varepsilon^2$, then we expand the vector function $\varphi(\varepsilon^2)$ in Eq.~(\ref{e29}) as a Taylor series at $\varepsilon=0$. Here we choose $a=0$ (i.e., $p_0=e^{-2iy}$) to simplify the expansion expression of $\varphi$ to have
\bee
\varphi(\varepsilon^2)=\varphi^{(0)}+\varphi^{(1)}\varepsilon^2+\varphi^{(2)}\varepsilon^4+\varphi^{(3)}\varepsilon^6+\ldots,   \label{e271}
 \ene
where
\bee\nonumber
\varphi^{(0)}=\left(\begin{array}{c} \phi^{(0)} \vspace{0.1in} \\ \psi^{(0)} \end{array}\right)
=\left(\begin{array}{c} \sqrt{2}e^{-iy} \vspace{0.1in}\\ i\sqrt{2}e^{iy} \end{array} \right),
\ene
\bee\nonumber
 \varphi^{(1)}\!=\!\left(\begin{array}{c}\phi^{(1)}\\ \psi^{(1)} \end{array}\right)
 =\left(\begin{array}{c} -\frac{i\sqrt{2}}{4}e^{-iy}(4 x^2-16 i x y-192 x t-16 y^2+384 i y t+2304 t^2+1+4 x-8 i y-96 t) \vspace{0.1in}\\ \frac{\sqrt{2}}{4}e^{iy}(4 x^2-16 i x y-192 x t-16 y^2+384 i y t+2304 t^2+1-4 x+8 i y+96 t)
\end{array} \right),
\ene
 $(\phi^{(i)},\psi^{(i)})^T (i=2,3)$ are listed in \textbf{Appendix A}, and $(\phi^{(j)},\psi^{(j)})^T (j=4,5,...)$ are omitted here.

In terms of the above-mentioned analysis and Theorem 2, we can determine ${B^{(N-1)}}=\frac{\Delta {B^{(N-1)}}}{\Delta_N}$. Finally, we can find the higher-order rational solitons and RW solutions of Eq.~(\ref{cmkdv}) in terms of $\widetilde{p}_{N-1}$  in the form
\bee
 \label{kps}
  \widetilde{u}_{N-1}=2|\widetilde{p}_{N-1}|^2=2|p_0+2B^{(N-1)}|^2,
 \ene
where $p_0$ is an initial solution of Eqs.~(\ref{nls}) and (\ref{cmkdv2}).

To understand the obtained exact solutions of Eq.~(\ref{cmkdv}) in Theorem 2 (or Theorem 3 with $n=1,\, m_1=N-1$), we will discuss the solution (\ref{kps}) from the following five cases with $N =1,2,3,4,5$. For $N=1$, it follows from Eq.~(\ref{kpsol-p}) that we only have the trivial constant solution.
In this step, we have $m_1=N-1=0$, that is to say, we do not use the derivatives of $T(\lambda_1)\varphi(\lambda_1)$ to determine $B^{(0)}$ such that we can not give a new solution.

\vspace{0.1in}
 {\bf Case I.}\,  For $N=2$, based on the generalized perturbation $(1,1)$-fold DT, we can derive the first-order RW solution (regular rational solution) of the KP equation (\ref{cmkdv})
\bee
  \widetilde{u}_1(x,y,t)=2\left|e^{i[ax+(a^2-2)y+(4a^3-24 a)t]}+2B^{(1)}\right|^2=\dfrac{F}{G},
 \label{u1}
 \ene
with 
\bee\nonumber\begin{array}{rl}
F=& 256 y^2 x^2+512 y^4+320 y^2+512 a^4 y^4-192 a x y+12288 a y t+256 a x^3 y-12288 y^2 t x+1024 a^2 y^4 \vspace{0.05in}\\ 
  & +147456 y^2 t^2+1769472 a^3 y t^3+147456 a y t^2 x-12288 a x^2 y t-442368 a^4t^3 x-73728 a^2x^2 t^2
     \vspace{0.05in}\\
  & -6912 a^4t^2-192 a^2y^2 +36864 a^4y^2 t^2+ 12288 a^3y^3 t +884736 a^2t^3x+1536 a^2 x^3 t+1024 a x y^3 \vspace{0.05in}\\
  & +32 x^4+27648 a^4 x^2 t^2+5308416 a^4t^4 -2304 a^3y  t-1152 a^2x t+110592 x^2 t^2-147456 a^3x t^2 y \vspace{0.05in}\\ 
  & -6144 a^2x y^2 t+10616832 t^4+18432 a^4 x t y^2+110592 a^5 x t^2 y+9216 a^3 x^2 t y-1769472 t^3 x \vspace{0.05in}\\ 
  & +2304 x t+663552 a^8t^4 -48 x^2-27648 t^2+768 a^2 x^2 y^2+147456 a^2y^2 t^2 +221184 a^6 x t^3\vspace{0.05in}\\  
  & +1024 a^3 x y^3+442368 a^7 y t^3+110592 a^6 y^2 t^2+12288 a^5 y^3 t+73728 a^2t^2 -3072 x^3 t+18,
  \end{array}
  \ene
\bee\nonumber
 G=(576 a^4t^2 +192 a^3y t+16 a^2y^2 +96 a^2x t +16 a x y+2304 t^2+4 x^2+16 y^2+1-192 x t)^2,\qquad\quad\qquad\quad
 \ene
which contains a free parameter $a$. In fact, we can also obtain the generalized RW solution from solution (\ref{u1}) with the parameter $a$ in terms of Proposition 1, which contains four free parameters $\alpha\not=0,\, \beta,\, \gamma$ and $a$.

It follows from the solution (\ref{u1}) that the maximum amplitude $\widetilde{u}_1=18$ occurs at $ x = 12t(a^2+2),\, y = -12ta$ and $t = t$, the minimum amplitude $\widetilde{u}_1=0$ is reached at $x = 12t(a^2+2)\pm\frac{\sqrt{3}}{2},\, y = -12ta$ and $t=t$, where $t$ can be chosen as arbitrary real number. We have $\widetilde{u}_1\rightarrow 2$ as $x, t\rightarrow \infty$.

In particular, when $a=0$, we have the regular rational solution of the KP equation (\ref{cmkdv})
\bee \label{u10}
 \widetilde{u}_{10}(x,y,t)=2\left\{1-\frac{8}{1+4(x-24t)^2+16y^2}+\frac{16(1+16y^2)}{[1+4(x-24t)^2+16y^2]^2}\right\},
\ene
Notice that though the symmetry $u(x,y,t)$ with $y=0$ reduces the KP Eq.~(\ref{cmkdv}) to the KdV equation with the external force $f(t)$ in the form
\begin{eqnarray}
\label{kdv1}
     u_t+6uu_x+u_{xxx}=f(t),
\end{eqnarray}
where $f(t)$ is a function of time, but the solution $\widetilde{u}_{10}(x,y,t)$ of KP equation (\ref{cmkdv}) with $y=0$ in the form
\bee \nonumber
 \widetilde{u}_{10}(x,y,t)\big|_{y=0}=\frac{32-16[1+4(x-24t)^2]}{[1+4(x-24t)^2]^2}+2
\ene
does not satisfy Eq.~(\ref{kdv1}) for any force $f(t)$, that is, we may not obtain the corresponding `rogue wave solutions' of the KdV equation by using the simple reduction $y=0$ of the RW solution of the KP equation. This case also appears in the following higher-order RW solutions.

According to Proposition 1, we can generate the generalized RW solutions from the solution (\ref{u10})
\begin{eqnarray}\label{u10g} \begin{array}{rl}
  \widetilde{u}_{10g}(x,y,t)=& \alpha^2\widetilde{u}_{10}(x,y,t)\Big|_{\{x\rightarrow\alpha x+\beta y+\gamma t,\,\, y\rightarrow\alpha^2y+6\alpha\beta t, \,\, t\rightarrow\alpha^3t\}}+\dfrac{3\beta^2-\alpha \gamma}{6\alpha^2} \vspace{0.1in}\\
  =& 2\left\{1-\dfrac{8}{1+4[\alpha x+\beta y+(\gamma-24\alpha^3)t]^2+16(\alpha^2y+6\alpha\beta t)^2}\right. \vspace{0.1in}\\ & \qquad \left.+\dfrac{16[1+16(\alpha^2y+6\alpha\beta t)^2]}{\{1+4[\alpha x+\beta y+(\gamma-24\alpha^3)t]^2+16(\alpha^2y+6\alpha\beta t)^2\}^2}\right\} +\dfrac{3\beta^2-\alpha \gamma}{6\alpha^2},
    \end{array}
 \end{eqnarray}
which contains three free parameters $\alpha\not=0,\,\beta$, and $\gamma$. When $\alpha=\mu/2,\, \beta=\nu\mu/2$, and
$\gamma=3\mu(\mu^2+\nu^2)/3$, the generalized rogue wave solution $\widetilde{u}_{10g}(x,y,t)$ given by Eq.~(\ref{u10g}) reduces to the known lump solution~\cite{kpm,kplum}
\bee
 u_{ls}=4\frac{-[x+\nu y+3(\nu^2-\mu^2)t]^2+\mu^2(y+6\nu t)^2+1/\mu^2}{\{[x+\nu y+3(\nu^2-\mu^2)t]^2+\mu^2(y+6\nu t)^2+1/\mu^2 \}^2}.
\ene
where $\nu$ and $\mu$ are real-valued constants, which approaches to zero as $x^2+y^2+t^2\rightarrow \infty$.

\begin{figure}
    \begin{center}
    {\scalebox{0.4}[0.4]{\includegraphics{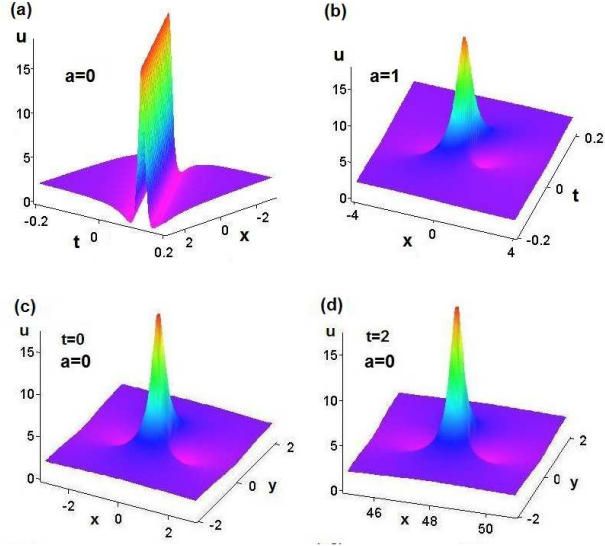}}}
    \end{center}
    \vspace{-0.15in} \caption{\small (color online).  The first-order rational soliton and RW solution $\widetilde{u}_{1}$ given by Eq.~(\ref{u1}) at different two-dimensional spaces. (a) $a=y=0$, (b) $a=1,\, y=0$, (c) $a=t=0$, (d) $a=0, t=2$.} \label{fig-rw1}
\end{figure}

\begin{figure}
	\begin{center}
	  {\scalebox{0.45}[0.45]{\includegraphics{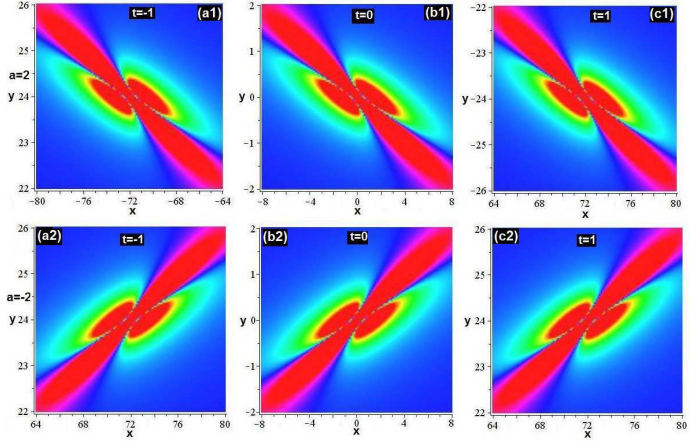}}}
	\end{center}
	\vspace{-0.15in} \caption{\small (color online).  The first-order rogue wave solution $\widetilde{u}_1$  given by Eq.~(\ref{u1}) with $a=2, -2$ at different times (a1)-(a2) $t=-1$; (b1)-(b2) $t=0$, (c1)-(c2) $t=1$. } \label{fig1-rw1-2}
\end{figure}

 The parameter $a$ plays a key role in the wave profile of solution (\ref{u1}) in $(x,t)$-space with $y=0$. For two cases $a=0$ and $a\not=0$, the solution (\ref{u1}) displays the different wave profiles in $(x,t)$-space with $y=0$ (see Fig.~\ref{fig-rw1}). When $a=0$, the solution (\ref{u1}) exhibits the W-shaped solitary wave, which is not localized (see Fig.~\ref{fig-rw1}a), but when $a\not=0$ (e.g., $a=1$), the parameter make the solution (\ref{u1}) generate the first-order RW profile (see Fig.~\ref{fig-rw1}b). That is to say, the parameter $a$ can modulate the solution (\ref{u1}) in the $(x,t)$-space with $y=0$ from the non-localized solution ($a=0$) to the localized solutions (first-order RW for $a=1$). For any parameter $a$, the solution (\ref{u1}) displays the same first-order RW profiles in both $(x,y)$-space with $t=0$ and $(y,t)$-space with $x=0$ (see Fig.~\ref{fig-rw1}).

In the following we mainly consider the localized wave profile of solution (\ref{u1}) in $(x,y)$-space for the fixed time: 
\begin{itemize}

\item {} When the parameter $a=0$, the shape of the first-order RW keeps covariant with $t$ changing; when time $t=0$, the core of the first-order RW is located at the origin; as time $t$ increases, the core moves along the positive $x$-axis; as time $t$ decreases, the core shifts along the negative $x$-axis and the first-order RW does not move in the $y$-direction, the corresponding wave profiles are shown in Figs.~\ref{fig-rw1}(c-d).

\item {} As the non-zero parameter $|a|$ becomes larger, the profile of $\widetilde{u}_1$ is becoming shrunk step by step in both $x$ and $y$ axes. If we fix the parameter $a\neq0$, then the first-order RW does not change its shape, and its core is always located at the origin with $t=0$. For the case $a>0$, as time $t$ increases, the first-order RW moves in the low right and upper left on the $(x,y)$-plane as time $t$ increases and decreases, respectively (see Figs.~\ref{fig1-rw1-2}(a1)-(c1)). For the case $a<0$, the first-order RW moves in the upper right and low left in the $(x,y)$ space with $t$ increasing and decreasing, respectively (see Figs.~\ref{fig1-rw1-2}(a2)-(c2)).
\end{itemize}

\begin{figure}[!t]
    \begin{center}
    {\scalebox{0.4}[0.4]{\includegraphics{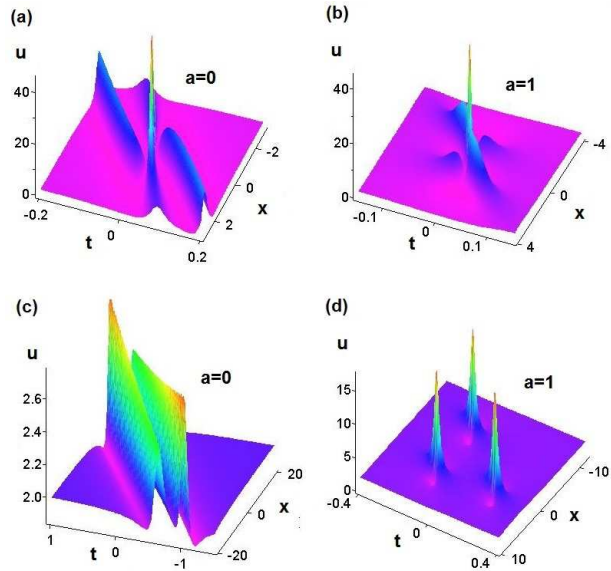}}}
    \end{center}
    \vspace{-0.15in} \caption{\small (color online).  The second-order rational soliton and RW solution $\widetilde{u}_2$ given by Eq.~(\ref{u2}) at different two-dimensional $(x,t)$ space with $y=0$.  (a) $a=b_1=c_1=0$ ($\widetilde{u}_2$ exhibits the strong interaction of two solitons), (b) $a=1,\, b_1=c_1=0$; (c) $a=c_1=0,\, b_1=100$, (d) $a=1,\, b_1=100,\, c_1=0$.} \label{fig-rw2-1}
\end{figure}

\begin{figure}[!t]
    \begin{center}
    {\scalebox{0.4}[0.4]{\includegraphics{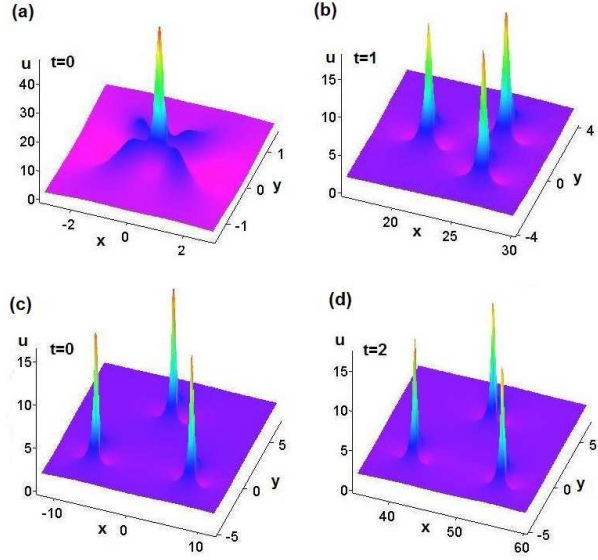}}}
    \end{center}
    \vspace{-0.15in} \caption{\small (color online).  The second-order RW solution $\widetilde{u}_2$ given by Eq.~(\ref{u2}) in $(x,y)$-space with at different time. (a) $a=b_1=c_1=t=0$, (b)  $a=b_1=c_1=0,\, t=1$, (c)   $a=c_1=t=0,\, b_1=300$, (d) $a=c_1=0,\, t=2,\, b_1=300$. } \label{fig4-rw2}
\end{figure}

\vspace{0.1in}
 {\bf Case II.}\,  For $N=3$, based on the generalized perturbation $(1,2)$-fold DT, we derive the second-order RW solution  of the KP equation (\ref{cmkdv})
\begin{eqnarray}
  \widetilde{u}_2(x,y,t)=2\left|e^{i[ax+(a^2-2)y+(4a^3-24 a)t]}+\frac{2\Delta {B^{(2)}}}{\Delta_3}\right|^2,  \label{u2}
 \end{eqnarray}
which contains three free parameters $a,\, b_1$ and $c_1$ and is omitted here because it is rather tedious.
According to Proposition 1, we can also generate the generalized second-order RW solution from the solution (\ref{u2}). Here we only consider the solution (\ref{u2}).

Since the parameters $b_1$ and $c_1$ have the similar effect on the solution (\ref{u2}), thus we have two parameters $a$ and $b_1$ with $c_1=0$.   For four cases of the parameters $(a=b_1=0)$, $(a\not=0,\ b_1=0)$, $(a=0,\, b_1\not=0$) and $(ab_1\not=0)$, the solution (\ref{u2}) displays the different wave profiles in $(x,t)$-space with $y=0$ (see Figs.~\ref{fig-rw2-1}):

 \begin{itemize}

 \item{} When $a=b_1=0$, solution (\ref{u2}) exhibits the elastic interaction of two soliton solutions, which is not localized (see Fig.~\ref{fig-rw2-1}a),

 \item{} When $a\not=0$ (e.g., $a=1$) and $b_1=0$, the solution (\ref{u2}) generates the strong interaction of two first-order RWs (see Fig.~\ref{fig-rw2-1}b), that is, the parameter $a$ can modulate the solution (\ref{u2}) in the $(x,t)$-space with $y=0$ from the non-localized solution ((see Fig.~\ref{fig-rw2-1}a) for $a=0,\, b_1=0$) into the localized second-order RW solutions (see Fig.~\ref{fig-rw2-1}b) for $a=1,\, b_1=0$).

  \item{} When $a=0$ and $b_1\not=0$ (e.g., $b_1=100$), the solution (\ref{u2}) is split into two soliton solutions without any interaction (i.e., two  parallel solitons), which is not localized. Moreover, the amplitude of one soliton becomes high and another one becomes low as $|x|,\, |t|$ increase (see Fig.~\ref{fig-rw2-1}c). Moreover, it follows from Fig.~\ref{fig-rw2-1}c that the width of the upper solitary wave becomes narrow and another one becomes wide as $|x|,\, |t|$ increase.

 \item{} When $a\not=0$ (e.g., $a=1$) and $b_1=100$, the solution (\ref{u2}) is split into three first-order RWs without  any interaction (see Fig.~\ref{fig-rw2-1}d). That is to say, the parameter $a$ can modulate the solution (\ref{u2}) in the $(x,t)$-space with $y=0$ from the non-localized solution ($a=0$) into localized solutions (three first-order RWs for $a=1$).

 \end{itemize}

Notice that it follows from Figs.~\ref{fig-rw2-1}(a) and (c) that for the fixed $a=0$, the solution (\ref{u2}) exhibits both non-localized wave profile in $(x,t)$-space with $y=0$ for the different parameter $b_1$. But the parameter $b_1$ can split the solution (\ref{u2}) into
two parallel solitons for $b_1=100$ (see Fig.~\ref{fig-rw2-1}c) from the strong interaction of two solitons for $b_1=0$ (see Fig.~\ref{fig-rw2-1}a). Moreover, it follows from Figs.~\ref{fig-rw2-1}(b) and (d) that for the fixed $a=1$, the solution (\ref{u2}) exhibits both localized wave profiles in $(x,t)$-space with $y=0$ for the different parameter $b_1$. But the parameter $b_1$ can split the solution (\ref{u2}) into three first-order rogue waves for $b_1=100$ (see Fig.~\ref{fig-rw2-1}d) from the strong interaction of second-order RW for $b_1=0$ (see Fig.~\ref{fig-rw2-1}b).

 For any parameters $a$ and $b_1$, the solution (\ref{u2}) displays the similar second-order RW profiles in both $(x,y)$-space with $t=0$ and $(y,t)$-space with $x=0$. Thus we only consider the solution (\ref{u2}) in $(x,y)$-space with the fixed time.

   Next, we study the wave profiles of the second-order RW solution with different parameters of $a=c_1=0,\, b_1$. It follows from Eq.~(\ref{u2}) that the maximum amplitude $\widetilde{u}_2=50$ occurs at the origin and the minimum value of $\widetilde{u}_2$ is $0$.  When $x\rightarrow \infty$, $y\rightarrow \infty$, $u\rightarrow 2$, and $a=0=b_1=c_1=0$, Figs.~\ref{fig4-rw2}(a) and (b) display the profiles of strong and weak interactions of second-order RW at $t=0$ and $t=1$, respectively. If we choose $b_1\not=0$ (e.g. $b_1=300)$ and $a=c_1=0$, then we have the profiles of weak interactions of second-order RWs at $t=0$ ( Fig.~\ref{fig4-rw2}c) and $t=2$ (Fig.~\ref{fig4-rw2}d), respectively, in which the second-order RW has been split into three first-order RWs that array an isosceles triangle structure. It follows from Figs.~\ref{fig4-rw2}(c) and (d) that the centers of triangle structures far away from the origin as $t$ increases.  By comparing
   Figs.~\ref{fig4-rw2}(a,b,c), we find that the parameter $b_1$ and $t$ can modulate the second-order RW into the separable three first-order RWs.

\begin{figure}[!t]
    \begin{center}
        {\scalebox{0.58}[0.5]{\includegraphics{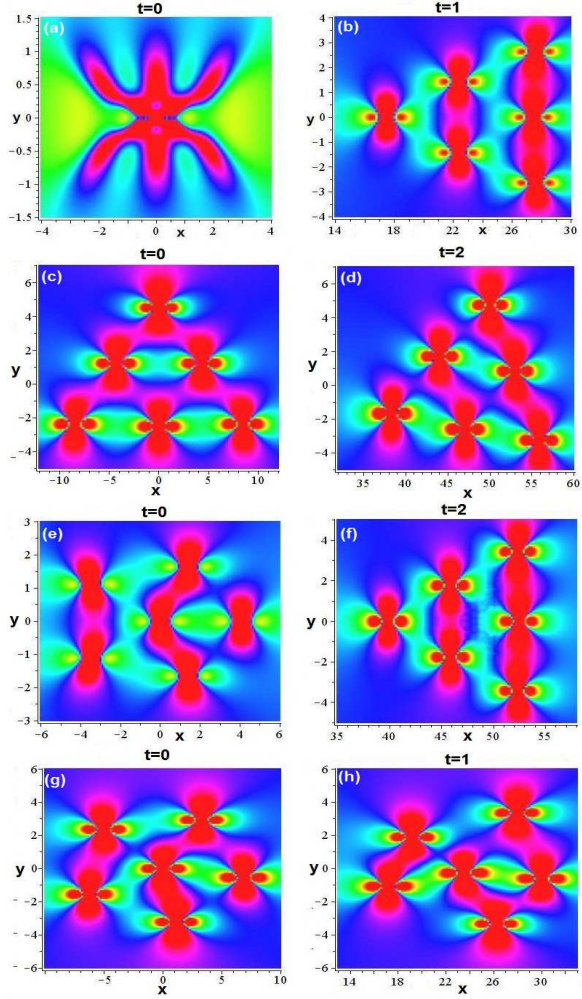}}}
    \end{center}
    \vspace{-0.15in} \caption{\small (color online).  The third-order RW solution $\widetilde{u}_3$ given by Eq.~(\ref{u3}) with $a=c_1=0$ at different time.
    (a) $b_1=b_2=c_2=t=0$, (a) $b_1=b_2=c_2=0,\, t=1$, (c) $b_1=100,\, b_2=c_2=t=0$, (d) $b_1=100,\, b_2=c_2=0,\ t=2$, (e) $b_2=100,\, b_1=c_2=t=0$, (f) $b_2=100,\, b_1=c_2=0,\ t=2$,  (g) $b_2=c_2=1000,\, b_1=t=0$, (h) $b_2=c_2=1000,\, b_1=0,\ t=1$. } \label{fig6-rw3}
\end{figure}

\vspace{0.1in}
 {\bf Case III.} \,  For $N=4$, based on the generalized perturbation $(1,3)$-fold DT, we can derive the third-order RW solution with five parameters $a,\,b_1,\,c_1,\,b_2,\,c_2$ as below:
\begin{eqnarray}
 \widetilde{u}_3(x,y,t)=2\left|e^{i[ax+(a^2-2)y+(4a^3-24 a)t]}+\frac{2\Delta B^{(3)}}{\Delta_4}\right|^2.   \label{u3}
 \end{eqnarray}
which is omitted here since it is so complicated. The effect of the parameter $a$ in solution (\ref{u3}) is similar to one in the first-order and second-order RWs. It follows from Eq.~(\ref{u3}) that the maximum amplitude $\widetilde{u}_3=98$ occurs at the origin and the minimum value of $\widetilde{u}_3$ is $0$. When $x\rightarrow \infty$ and $y\rightarrow \infty$, we have $\widetilde{u}_3\rightarrow 2$. In the following we discuss some special structure of the third-order RW solution (\ref{u3}):

\begin{itemize}

  \item{} \, When the parameters $a=b_{1,2}=c_{1,2}=0$, the wave structures of the third-order RW are shown in Figs.~\ref{fig6-rw3}(a,b). When time $t$ increases,
  the third-order RW (see Fig.~\ref{fig6-rw3}a for $t=0$) is split into  the six first-order RWs and array a triangle structure (see Fig.~\ref{fig6-rw3}b for $t=1$).

\item{} \, When the parameters $a=b_2=c_{1,2}=0$ and $b_1=100$, the wave structures of the third-order RW solution (\ref{u3}) are displayed in Figs.~\ref{fig6-rw3}(c,d). The third-order RW is made up of the six first-order RWs, which array a regular triangle structure regardless of the value of $t$, but the shape of the triangle may be changed as time increases or decreases.

\item{} \, When the parameters $a=b_1=c_{1,2}=0$ and $b_2=100$, the wave structures of the third-order RW are shown in Figs.~\ref{fig6-rw3}. For $t=0$, the third-order RW is made up of the six first-order RWs, which almost array array a regular pentagon (one sits in the center, while the rest ones are located on the vertices of the pentagon, see Fig.~\ref{fig6-rw3}e). The six first-order RWs evolve a regular triangle structure again with time increasing (see Fig.~\ref{fig6-rw3}f).

\item{} \, When the parameters $a=b_1=c_1=0$ and $b_2=c_2=1000$, the third-order RW is split into six first-order RWs at $t=0$, which array a regular pentagon (one sits in the center, while the rest ones are located on the vertices of the pentagon, see Fig.~\ref{fig6-rw3}g). When $t\neq0$ (e.g., $t=1$), the six first-order RWs array an irregular pentagon (see Fig.~\ref{fig6-rw3}h).

\end{itemize}

 {\bf Case IV.} \,  For $N=5$, we have $B^{(4)}=\frac{\Delta {B^{(4)}}}{\Delta_5}$ from Eq.~(\ref{kpsysg}) such that we obtain the fourth-order RW solution of the KP equation (\ref{cmkdv}) in the form
 \bee
 \label{solu4}
  \widetilde{u}_4(x,y,t)=2\left|e^{i[ax+(a^2-2)y+(4a^3-24 a)t]}+\frac{2\Delta B^{(4)}}{\Delta_5}\right|^2,
   \ene
   which contains seven parameters $a,\, b_{1,2,3,}\, c_{1,2,3}$ and is omitted here since it is so complicated. The parameters $b_{1,2,3,}\, c_{1,2,3}$ can modulate the abundant structures of the fourth-order RW (see Fig.~\ref{fig10-rw4}) for four cases:

\begin{figure*}[!t]
    \begin{center}
        {\scalebox{0.75}[0.75]{\includegraphics{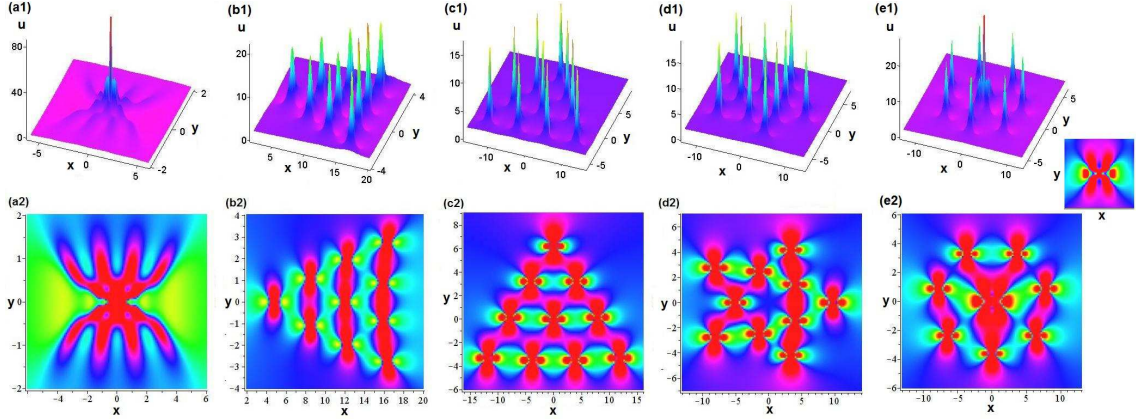}}}
    \end{center}
    \vspace{-0.15in} \caption{\small (color online).  The fourth-order RW solution $\widetilde{u}_4$ given by Eq.~(\ref{solu4})  with $a=c_{1,2,3}=0$ at different time and parameters (a1)-(a2)  $t=0,\, b_{1,2,3}=0$; (b1)-(b2) $t=0.5,\, b_{1,2,3}=0$, (c1)-(c2) $b_1=100, t=b_{2,3}=0$, (d1)-(d2) $t=0,\, b_{1,3}=0,\, b_2=1000$, (e1)-(e2) $t=0,\, b_{1,2}=0,\, b_3=10000$. The small figure displays the center part of Fig.~\ref{fig10-rw4}(e2), which seems to be a second-order RW.} \label{fig10-rw4}
\end{figure*}

\begin{itemize}

  \item{} \, When $a=t=0,\, b_{1,2,3}=c_{1,2,3}=0$, the fourth-order RW solution stays beside the origin in $(x,y)$-plane, that is, the four waves generate the interaction at the point $(x,y)=(0,0)$ (see Figs.~\ref{fig10-rw4}(a1)-(a2)).

 \item{} \, When at least one in these parameters $\{b_{1,2,3,}\, c_{1,2,3}\}$ is not zero, the fourth-order RW can be spit into several first-order and second-order RWs. In fact, the effects of $c_i$ are similar to one of $c_i$ for every index $i$. Here we fix $c_{1,2,3}=0$. For example, when $b_1\gg 0$ (e.g., $b_1=100$) and $b_{2,3}=0$, the split ten first-order RWs array an isosceles triangle, which seems to be a `ten-pin bowling' for $t=0, 0.5$ (see Figs.~\ref{fig10-rw4}(b1)-(b2) and (c1)-(c2)).

 \item{} \, When  $b_2 \gg 0$ (e.g., $b_2=10^3$) and $b_{1,3}=0$, the split ten first-order RWs array two different pentagons which seems to admit the same center point for $t=0$ (see Figs.~\ref{fig10-rw4}(d1)-(d2)).

 \item{} \, When  $b_3 \gg 0$ (e.g., $b_3=10^4$) and $b_{1,2}=0$, the splitted first-order rogue waves array a heptagon with the center being a second-order RW (see Figs.~\ref{fig10-rw4}(e1)-(e2)).

\end{itemize}

For the other cases $N>5$, we can also find the higher-order rational solitons and RW solutions of Eq.~(\ref{cmkdv}), which also possess the abundant structures.

\section{The higher-order rational solitons and RWs of the (2+1)-D generalized KP equation}

In the section, we will consider the (2+1)-dimensional gKP equation (\ref{gkp}).  When $\kappa=-\frac{1}{4}$, Eq.~(\ref{gkp}) becomes the (2+1)-dimensional KP equation, which has only different coefficients with Eq.~(\ref{cmkdv}). Some soliton solutions of Eq.~(\ref{gkp}) have been obtained by using the usual $N$-fold DT~\cite{you}. In the following, we present the generalized perturbation $(n, M)$-fold Darboux transformation for Eq.~(\ref{gkp}) such that we give new multi-RW solutions.
Eq.~(\ref{gkp}) can be separated into the focusing NLS equation
\begin{eqnarray}
 i p_{y} -p_{xx}-\frac{1}{2}|p|^2p=0 \label{nls22}
\end{eqnarray}
and the (1+1)-dimensional complex mKdV equation
\begin{eqnarray}
  p_{t}+p_{xxx}+\frac{3}{2}|p|^2p_x=0 \label{cmkdv22}
\end{eqnarray}
under the following constraint
\begin{eqnarray}
  u=\frac{1}{2}|p|^2, \label{yueshu1}
\end{eqnarray}
where $p=p(x,y,t)$ is the complex function of the variables $x,y,t$. If $p(x,y,t)$ solves Eqs.~(\ref{nls22}) and~(\ref{cmkdv22}), then the corresponding solution of the gKP equation (\ref{gkp}) can be generated in terms of the constraint (\ref{yueshu1}).

\subsection{The generalized perturbation DT}

The Lax representations of Eqs.~(\ref{nls22}) and~(\ref{cmkdv22}) are given as follows:
\begin{eqnarray}
\hspace{-0.9in}\varphi_x=U \varphi, \quad  U=\frac{i}{2}\left(\begin{array}{cc}\lambda & p \vspace{0.1in} \\  p^{*}  & -\lambda \end{array}\right),  \label{lax11}
 \qquad
\end{eqnarray}
\begin{eqnarray}
 \varphi_{y}=V \varphi,\quad V\!=\!\frac{i}{4}\left(\!\begin{array}{cc}
        2\lambda^2-|p|^2            & 2\lambda p -2i p_x  \vspace{0.1in} \\
        2\lambda p^{*}+2ip^{*}_x      & -2\lambda^2+|p|^2 \end{array} \!\right), \,\,\, \label{lax21}
\end{eqnarray}
\bee \label{lax31}
 \hspace{-0.9in} \varphi_{t}=W \varphi, \quad W=\left(\begin{array}{cc} W_{11} & W_{12} \vspace{0.1in}\cr W_{21} & W_{22}\end{array}\right) \quad
 \ene
with
 \bee \nonumber \begin{array}{l}
W_{11}=\frac{1}{2} i \lambda^3-\frac{1}{4} i \lambda |p|^2+\frac{1}{4}(pp^{*}_x-p_xp^{*}), \,\
W_{12}= \frac{1}{2} i \lambda^2p +\frac{1}{2}\lambda p_x -\frac{1}{2}i( p_{xx}+\frac{1}{2}|p|^2p), \vspace{0.1in} \\
W_{21}=\frac{1}{2} i\lambda^2 p^{*}-\frac{1}{2}\lambda p^{*}_x-\frac{1}{2}i(p^{*}_{xx}+\frac{1}{2}|p|^2p^{*}),\,\,
W_{22}=-\frac{1}{2} i \lambda^3+\frac{1}{4} i \lambda |p|^2-\frac{1}{4}(pp^{*}_x-p_xp^{*}).
\end{array}\ene
It is easy to show that two zero curvature equations $U_y-V_x+[U, V]=0$ and $U_t-W_x+[U, W]=0$ yield Eqs.~(\ref{nls22}) and~(\ref{cmkdv22}), respectively.

Similar to the steps for the generalized DTs of the KP equation (\ref{cmkdv}) in Sec. 2, we can obtain the following theorem:\\

\noindent {\bf Theorem 4.}\, {\it Let $\varphi_i(\lambda_i)=(\phi_i(\lambda_i),\psi_i(\lambda_i))^T\,\, (i=1,2,...,n)$ be column vector solutions of the spectral problem (\ref{lax11})-(\ref{lax31}) for the $n$ spectral parameters $\lambda_1, \lambda_2,...,\lambda_n$ and initial solution $p_0$ of Eq.~(\ref{gkp}), respectively, then the generalized perturbation $(n, M)$-fold DT of Eq.~(\ref{gkp}) is given by
\bee\label{kpsolg2}
 \widetilde{u}_{N-1}=\frac{1}{2}|\widetilde{p}_{N-1}|^2=\frac{1}{2}\left|p_0-2B^{(N-1)}\right|^2,
\ene
where $B^{(N-1)}=\Delta B^{(N-1)}/\Delta_N^{(n)}$  with
$\Delta_N^{(n)}={\rm det}([\Delta^{(1)}...\Delta^{(n)}]^T)$ and $\Delta^{(i)}\!=\!(\Delta^{(i)}_{j,s})_{2N\times 2N}=$
\bee\nonumber
\left[\begin{array}{llllllll}
{\lambda_i}^{N-1} {\phi_i^{(0)}} & {\lambda_i}^{N-2} {\phi_i^{(0)}}  &\ldots & {\phi_i^{(0)}} & {\lambda_i}^{N-1} {\psi_i^{(0)}} & {\lambda_i}^{N-2} {\psi_i^{(0)}}  &\ldots & {\psi_i^{(0)}} \vspace{0.1in} \\
\Delta_{2,1}^{(i)}& \Delta_{2,2}^{(i)}  &\ldots &{\phi_i^{(1)}}  & \Delta_{2,N+1}^{(i)} &  \Delta_{2,N+2}^{(i)} &\ldots &{\psi_i^{(1)}} \vspace{0.1in}\\
 \ldots & \ldots      &\ldots           &\ldots &\ldots & \ldots      &\ldots           &\ldots  \vspace{0.1in} \\
\Delta_{m_i+1,1}^{(i)}  & \Delta_{m_i+1,2}^{(i)} &\ldots &{\phi_i^{(m_i)}} &\Delta_{m_i+1,N+1}^{(i)}  & \Delta_{m_i+1,N+2}^{(i)}  &\ldots &{\psi_i^{(m_i)}} \vspace{0.1in}\\
{\lambda_i^*}^{(N-1)} {\psi_i^{(0)}}^* & {\lambda_i^*}^{(N-2)} {\psi_i^{(0)}}^*  &\ldots & {\psi_i^{(0)}}^* & -{\lambda_i^*}^{(N-1)} {\phi_i^{(0)}}^* & -{\lambda_i^*}^{(N-2)} {\phi_i^{(0)}}^*  &\ldots & -{\phi_i^{(0)}}^*\vspace{0.1in} \\
\Delta_{m_i+3,1}^{(i)} & \Delta_{m_i+3,2}^{(i)}  &\ldots &{\psi_i^{(1)}}^*  &\Delta_{m_i+3,N+1}^{(i)} & \Delta_{m_i+3,N+2}^{(i)}  &\ldots &-{\phi_i^{(1)}}^* \vspace{0.1in}\\
 \ldots & \ldots      &\ldots           &\ldots &\ldots & \ldots      &\ldots           &\ldots  \\
\Delta_{2(m_i+1),1}^{(i)}  & \Delta_{2(m_i+1),2}^{(i)}  &\ldots &{\psi_i^{(m_i)}}^* &\Delta_{2(m_i+1),N+1}^{(i)} & \Delta_{2(m_i+1),N+2}^{(i)}  &\ldots &-{\phi_i^{(m_i)}}^* \\
\end{array}\right],
\ene
where $\Delta_{j,s}^{(i)}\, (1\leq j, s\leq 2N$) are given by the following formulae:
\bee\nonumber
\Delta_{j,s}^{(i)}=\begin{cases}
 \sum\limits_{k=0}^{j-1}C^{k}_{N-s} {\lambda_i}^{N-s-k} {\phi_i^{(j-1-k)}}
\quad {\rm for} \quad 1\leq j\leq m_i+1,\, 1\leq s\leq N, \vspace{0.1in} \\
\sum\limits_{k=0}^{j-1}C^{k}_{2N-s} {\lambda_i}^{2N-s-k} {\psi_i^{(j-1-k)}}\quad {\rm for} \quad 1\leq j\leq m_i+1,\, N+1\leq s\leq 2N, \vspace{0.1in}\\
\sum\limits_{k=0}^{j-(N+1)}C^{k}_{N-s} {\lambda_i}^{*(N-s-k)} {\psi_i^{(j-N-1-k)*}} \quad {\rm for} \quad
m_i+2\leq j\leq 2(m_i+1),\, 1\leq s\leq N, \vspace{0.1in}\\
-\sum\limits_{k=0}^{j-(N+1)}C^{k}_{2N-s} {\lambda_i}^{*(2N-s-k)} {\phi_i^{(j-N-1-k)*}}\quad {\rm for} \quad
m_i+2\leq j\leq 2(m_i+1),\, N+1\leq s\leq 2N \end{cases} \ene
and $\Delta {B^{(N-1)}}$ is found from the determinant $\Delta_N^{(n)}$ by replacing its
$(N+1)$-th column with the  vector $(b^{(1)}\cdots b^{(n)})^T$, where $b^{(i)}=(b_j^{(i)})_{2(m_i+1)\times 1}$ with
\bee\nonumber
b_j^{(i)}=\begin{cases}
-\sum\limits_{k=0}^{j-1}C^{k}_{N} {\lambda_i}^{N-k} {\phi_i^{(j-1-k)}} \quad {\rm for} \quad  1\leq j\leq m_i+1, \vspace{0.1in}  \\ -\sum\limits_{k=0}^{j-(N+1)}C^{k}_{N} {\lambda_i}^{*(N-k)} {\psi_i^{(j-N-1-k)*}} \quad {\rm for} \quad m_i+2\leq j\leq 2(m_i+1).
\end{cases} \ene
}

From Theorem 4, we can give some higher-order rational solitons and RW solutions of Eq.~(\ref{gkp}) in terms of determinants using the generalized perturbation $(n, M)$-fold DT.

\subsection{The higher-order rational solitons and RW solutions}

We here consider only one spectral parameter $\lambda=\lambda_1$, i.e., $n=1,\, m_1=N-1$ in Theorem 4. We begin with the non-trivial `seed' plane-wave solution of Eqs.~(\ref{nls22}) and (\ref{cmkdv22})
 \bee
  p_0(x,y,t)=c e^{-i[ax+(-a^2+\frac{1}{2}c^2)y+(a^3-\frac{3}{2} a c^2)t]},
   \ene
where $a,\, c$ are the real parameters, the wave numbers in $x-$ and $y-$directions are $-a$ and $a^2-\frac{1}{2}c^2$, respectively. It is known that the phase velocities in $x-$ and $y-$directions are $a^2-\frac{3}{2} c^2$ and $\frac{a^3-\frac{3}{2} a c^2}{-a^2+\frac{1}{2}c^2}$, respectively, and the group velocities in $x-$ and $y-$directions are $\frac{3}{2} c^2-3a^2$ and $\frac{3}{4} c^2-\frac{3}{2}a^2$, respectively.

Substitute $p_0(x,y,t)$ into the linear spectral problem (\ref{lax11})-(\ref{lax31}). As a result, we can give the
eigenfunction solution of the Lax pair (\ref{lax11})-(\ref{lax31}) as
follows:
\bee \nonumber
\varphi=\left[\begin{array}{c}(-C_1 e^{-A}+C_2
e^{A})e^{-\frac{i}{2}[ax+(\frac{c^2}{2}-a^2)y+(a^3-\frac{3}{2}a c^2)t]} \vspace{0.1in}\\
(C_2 e^{-A}+C_1 e^{A})e^{\frac{i}{2}[ax+(\frac{c^2}{2}-a^2)y+(a^3-\frac{3}{2}a c^2)t]}
\end{array} \right], \\ \label{e29x} \ene with
\bee\nonumber
\begin{array}{l}
C_1=C_-,\,\, C_2=C_+, \,\, C_{\pm}=\sqrt{\dfrac{\sqrt{c^2+a^2+\lambda^2+2a\lambda}\pm (a+\lambda)}{c}}, \vspace{0.1in}\\
A=i\sqrt{c^2+a^2+\lambda^2+2a\lambda}\left[\frac{1}{2}x+\frac{1}{2}(\lambda- a)y+\frac{1}{2}(4\lambda^2+a^2-a\lambda-\frac{1}{2}c^2)t+\Theta(\varepsilon)\right], \,\,
  \Theta(\varepsilon)=\sum\limits_{k=1}^{N}(b_k+ic_k)\varepsilon^{2k},
\end{array} \ene
where $b_k,\, c_k (k=1,2,...,N)$ are real parameters, and $\varepsilon$ is a small parameter.

Next, we fix the eigenvalue $\lambda_{1}=-a+i c$, and set
$\lambda=\lambda_{1}+\varepsilon^2$, then expand the vector function
$\varphi(\varepsilon^2)$ in Eq.~(\ref{e29x}) as a Taylor series at
$\varepsilon=0$. Here we choose $a=0,\, c=1$ (i.e., $p_0=e^{-iy/2}$) to
simplify the expansion expression of $\varphi$ to get
 \bee
\varphi(\varepsilon^2)=\varphi^{(0)}+\varphi^{(1)}\varepsilon^2+\varphi^{(2)}\varepsilon^4+\varphi^{(3)}\varepsilon^6+\ldots,
\nonumber
 \ene
where $(\phi^{(i)},\psi^{(i)})^T (i=0,1,2,3,...,)$ are omitted. Here we only give the first-order rational soliton and  RW solution of Eq.~(\ref{gkp}) with $a=0$ and $c=1$ as below:
\bee \label{u20}
 \widetilde{u}_{20}(x,y,t)=\frac{16[4-(2x-3t)^2+4y^2]}{[4+(2x-3t)^2+4y^2]^2}+\frac{1}{2}.
\ene

Similarly, we can also give the higher-order rational solitons and RW solutions of Eq.~(\ref{gkp}) by using Theorem 4.

\section{Conclusions and discussions}

In conclusion, we have presented a novel approach to construct the generalized perturbation $(n, M)$-fold DT for both the (2+1)-D KP equation (\ref{cmkdv}) and the gKP equation (\ref{gkp}) such that their higher-order rational solitons and RW solutions are found. The constructive procedure is divided into two steps: Firstly, a brief introduction of the usual $N$-fold DT in matrix form for Eqs.~(\ref{cmkdv}) and (\ref{gkp}) are given. Secondly a detailed derivation of the generalized perturbation $(n, M)$-fold DTs for Eqs.~(\ref{cmkdv}) and (\ref{gkp}) are discussed in terms of the $N$-order Darboux matrix, the Taylor expansion and a limit procedure. Finally, the generalized perturbation $(1, N-1)$-fold DT allows us to obtain higher-order RW solutions of Eqs.~(\ref{cmkdv}) and (\ref{gkp}) in terms of determinants in a unified way.
These spatial-temporal structures of higher-order rational solitons and RWs may need further research due to more parameters. We hope that our results are useful for understanding the generation mechanism and finding possible application of RWs. We believe that the used idea is rather general and could be applied to other physically interesting nonlinear wave models as well. By comparing our results with the known results for Eqs.~(\ref{cmkdv}) and (\ref{gkp}), our main achievements are listed as below:

\begin{itemize}

\item{}\, We have given in detail the generalized perturbation $(n, M)$-fold DTs for  Eqs.~(\ref{cmkdv}) and (\ref{gkp}), and directly obtained their higher-order rational solitons and RW solutions in terms of determinants.

\item{} \, Compared with the usual $N$-fold DT method, the generalized perturbation DT method can give the higher-order RW solutions in terms of determinants with the same one eigenvalue, while the usual $N$-fold DT method can only generate solitons with $N$ distinct eigenvalues. Our generalized perturbation $(n, M)$-fold DT method and results for the (2+1)-dimensional KP equation (\ref{cmkdv}) and the gKP equation (\ref{gkp}) contain both the known results and new ones.

\item {} \, We use the generalized perturbation $(1, N-1)$-fold DT with only one spectral parameter to study multi-RW solutions of Eqs.~(\ref{cmkdv}) and (\ref{gkp}). In fact, we can also use more than one spectral parameters to study their  generalized perturbation $(n, M)$-fold DTs, e.g., $n=2$ and $m_2=0$ and $m_1=N-2$ such that we may obtain their interactions of RW solutions and multi-soliton solutions, which will be studied in future  investigations.
    \end{itemize}

\noindent {\bf Acknowledgements}

\vspace{0.05in}
The authors thank the referees for their valuable suggestions and comments.
This work was partially supported by the Beijing Natural Science Foundation under Grant No. 1153004, China Postdoctoral Science Foundation under Grant No. 2015M570161,
the NSFC under Grant Nos. 61471406 and 11571346, and the Youth Innovation Promotion Association CAS.
\vspace{0.1in}

{\baselineskip=15pt \noindent\textbf{Appendix A.} \vspace{0.1in} \\ 
\noindent $\phi^{(2)}= - \frac{\sqrt{2}}{96}e^{-iy}(1769472 i  y t^3+ 512 i  x y^3+24 x-3648 t- 672 i  x y- 12288 i  y^3 t- 110592 i  y t^2+ 9216 i  x y t- 192 i  x^2 y- 128 i  x^3 y+ 9216 i  x^2 y t-96 c_1- 221184 i  x y t^2+72 x^2-9600 x t+188928 t^2-2304 x^2 t+55296 x t^2+32 x^3-442368 t^3-384 x y^2+9216 y^2 t-1056 y^2+16 x^4+5308416 t^4-192 c_1 x+4608 c_1 t-1536 x^3 t+55296 x^2 t^2-884736 x t^3+18432 x y^2 t-384 x^2 y^2-221184 y^2 t^2+256 y^4+384 y b_1- 4608 i  t b_1+ 192 i  x b_1+ 384 i  c_1 y- 240 i  y+ 96 i  b_1+ 28416 i  y t+ 256 i  y^3-3),$\vspace{0.1in}\\
$\psi^{(2)}= -\frac{\sqrt{2}}{96}e^{iy}( 256 i  y^4+ 5308416 i  t^4+ 16 i  x^4+ 384 i  y b_1+ 188928 i  t^2+ 72 i  x^2- 884736 i  x t^3-9216 x^2 y t- 24 i  x+ 3648 i  t+ 96 i  c_1- 1536 i  x^3 t- 192 i  c_1 x+ 4608 i  c_1 t- 1056 i  y^2+ 442368 i  t^3- 32 i  x^3-240 y+221184 x y t^2+9216 x y t+96 b_1+ 18432 i  x y^2 t-110592 y t^2-192 x^2 y+256 y^3+672 x y-192 x b_1-28416 t y+4608 t b_1-1769472 y t^3+128 x^3 y-512 y^3 x+12288 y^3 t-384 y c_1- 384 i  x^2 y^2- 9600 i  x t+ 2304 i  x^2 t- 9216 i  y^2 t+ 55296 i  x^2 t^2+ 384 i  x y^2- 221184 i  y^2 t^2-3 i- 55296 i  x t^2),$\vspace{0.1in}\\
$\phi^{(3)}= \frac{\sqrt{2}}{5760}e^{-iy}( 8847360 i  y^4 t^2- 7680 i  x^3 y^2-1981440 x^2 y t- 26542080 i  x^2 t^3+ 17268480 i  t^2+6635520 y t^2 x^2- 23040 i  y c_2+ 1105920 i  x^3 t^2+ 318504960 i  t^4 x+ 540 i  x^2+ 192 i  x^5- 5760 i  b_2+737280 x y^3 t- 59760 i  y^2+46080 x y^2 b_1+ 15360 i  y^4 x- 1105920 i  x t b_1 y-23040 x y c_1- 180 i  x- 133920 i  t- 1440 i  c_1+ 80640 i  y^4+2520 y+ 2096824320 i  t^4+ 1200 i  x^4+ 5068800 i  x y^2 t- 3317760 i  t^2 c_1+ 106168320 i  y^2 t^3- 5760 i  x^2 c_1- 6635520 i  x t^2 c_1+ 276480 i  x^2 t c_1+ 368640 i  x^3 y^2 t- 13271040 i  y^2 t^2 x^2+ 212336640 i  y^2 t^3 x+65249280 x y t^2-184320 x^3 y t- 23040 i  x^4 t- 368640 i  y^4 t+ 23040 i  y^2 c_1- 1105920 i  y^2 t c_1+ 11335680 i  x t^2- 288000 i  x^2 t+ 13271040 i  t^2 b_1 y- 3840 i  x^3 c_1+ 46080 i  x y^2 c_1+ 2257920 i  y^2 t+ 23040 i  x y b_1+ 23040 i  x^2 b_1 y- 8640 i  c_1 x- 486720 i  x t+ 40320 i  y b_1- 737280 i  x y^4 t- 78520320 i  y^2 t^2- 74880 i  x^2 y^2+ 53084160 i  t^3 c_1- 63360 i  x y^2+ 12994560 i  x^2 t^2- 552960 i  y t b_1- 238080 i  x^3 t+ 576000 i  c_1 t- 1274019840 i  y^2 t^4- 13271040 i  x y^2 t^2- 3840 i  x^4 y^2-1704960 x y t- 30720 i  y^3 b_1+ 276480 i  x t c_1- 278691840 i  x t^3- 9216 i  x^5 t- 11520 i  x b_2+ 15360 i  y^4 x^2+ 552960 i  x^2 y^2 t+ 276480 i  t b_2+5760 c_2-1440 b_1- 17694720 i  x^3 t^3- 3057647616 i  t^5 x+ 318504960 i  t^4 x^2+ 552960 i  x^4 t^2+ 1440 i  x^3+29306880 y t^2+20160 x^2 y-57600 y^3+19440 x y-8640 x b_1-2125440 t y+576000 t b_1-663552000 y t^3+17280 x^3 y-130560 y^3 x+4116480 y^3 t-40320 y c_1-5760 x^2 b_1-3317760 t^2 b_1+276480 x t b_1-106168320 y t^3 x+552960 y t c_1-8847360 y^3 t^2+1920 x^4 y-15360 x^2 y^3+6144 y^5+23040 y^2 b_1+637009920 y t^4+11520 c_1 b_1+276480 x^2 t b_1-6635520 x t^2 b_1-3840 x^3 b_1+53084160 t^3 b_1-1105920 y^2 t b_1+1105920 x y c_1 t+11520 c_2 x-276480 c_2 t+4423680 y t^2 x^3-106168320 y t^3 x^2+737280 x^2 y^3 t-17694720 x y^3 t^2-23040 x^2 y c_1-92160 x^4 y t+1274019840 y t^4 x-13271040 y t^2 c_1+141557760 y^3 t^3+768 x^5 y-10240 x^3 y^3+12288 y^5 x-294912 y^5 t-6115295232 y t^5-23040 y b_2+30720 y^3 c_1+45 i- 4096 i  y^6- 126074880 i  t^3- 1528823808 i  t^5+ 64 i  x^6+ 5760 i  c_1^2+ 12230590464 i  t^6- 5760 i  b_1^2),$\vspace{0.1in}\\
$\psi^{(3)}= -\frac{\sqrt{2}}{5760}e^{iy}(45+ 2520 i  y+ 552960 i  c_1 y t- 17280 i  x^3 y+180 x+133920 t- 1105920 i  x y c_1 t+5760 c_1^2-17694720 x^3 t^3-23040 y^2 c_1+13271040 x y^2 t^2+ 3840 i  x^3 b_1+ 92160 i  x^4 y t+212336640 y^2 t^3 x-3057647616 t^5 x-552960 x^2 y^2 t+53084160 t^3 c_1- 106168320 i  x y t^3+ 276480 i  x t b_1+5760 b_2+1440 c_1+ 29306880 i  y t^2+ 663552000 i  t^3 y-13271040 y^2 t^2 x^2+368640 x^3 y^2 t+318504960 t^4 x^2+552960 x^4 t^2- 46080 i  x y^2 b_1-3840 x^3 c_1-9216 x^5 t+540 x^2-486720 x t+17268480 t^2+288000 x^2 t-11335680 x t^2-1440 x^3+126074880 t^3-11520 x b_2+63360 x y^2-2257920 y^2 t+276480 x^2 t c_1-6635520 x t^2 c_1-1105920 y^2 t c_1-59760 y^2+1200 x^4+2096824320 t^4-8640 c_1 x+576000 c_1 t-238080 x^3 t+12994560 x^2 t^2-278691840 x t^3+5068800 x y^2 t-74880 x^2 y^2-78520320 y^2 t^2+80640 y^4+40320 y b_1+15360 y^4 x^2- 11520 i  c_1 b_1+23040 x^4 t-318504960 t^4 x+5760 c_1 x^2+3317760 c_1 t^2-1105920 x^3 t^2+26542080 x^2 t^3+7680 x^3 y^2-106168320 y^2 t^3-276480 c_1 x t-192 x^5+1528823808 t^5-23040 y b_1 x+552960 y b_1 t-15360 y^4 x+368640 y^4 t+8847360 y^4 t^2+ 8640 i  x b_1-1274019840 y^2 t^4- 768 i  x^5 y+276480 t b_2- 57600 i  y^3-737280 x y^4 t+23040 x^2 b_1 y-1105920 x t b_1 y-3840 x^4 y^2+46080 x y^2 c_1+13271040 t^2 b_1 y- 11520 i  c_2 x- 141557760 i  y^3 t^3- 53084160 i  t^3 b_1-5760 b_1^2+64 x^6+12230590464 t^6-30720 y^3 b_1+ 1105920 i  y^2 t b_1+ 1920 i  x^4 y-23040 y c_2+ 23040 i  y^2 b_1- 276480 i  x^2 t b_1-4096 y^6+ 6144 i  y^5- 4116480 i  y^3 t+ 2125440 i  y t- 8847360 i  y^3 t^2+ 106168320 i  y t^3 x^2+ 276480 i  c_2 t- 1274019840 i  y t^4 x- 184320 i  x^3 y t- 576000 i  t b_1+ 23040 i  x^2 y c_1- 737280 i  x^2 y^3 t+ 17694720 i  x y^3 t^2- 23040 i  c_1 y x- 30720 i  y^3 c_1+ 1981440 i  x^2 t y+ 130560 i  x y^3- 5760 i  x^2 b_1+ 6115295232 i  y t^5- 3317760 i  t^2 b_1+ 5760 i  c_2+ 6635520 i  x t^2 b_1- 15360 i  x^2 y^3+ 637009920 i  t^4 y+ 6635520 i  x^2 y t^2- 65249280 i  x t^2 y+ 23040 i  y b_2+ 10240 i  x^3 y^3- 12288 i  y^5 x- 1440 i  b_1- 19440 i  x y+ 13271040 i  y t^2 c_1+ 20160 i  x^2 y+ 294912 i  y^5 t- 1704960 i  x y t- 4423680 i  y t^2 x^3+ 40320 i  y c_1+ 737280 i  x y^3 t).$ }

\end{document}